# Modern trends in Superconductivity and Superfluidity

Selected Chapters

*M.Yu. Kagan*

*P.L. Kapitza Institute for Physical Problems*

kagan@kapitza.ras.ru

In preparation for Springer-Verlag

Chapter 11. Fermionic superfluidity in three- and two-dimensional solutions of $^3$He in $^4$He.

11.1. Bardeen-Baym-Pines theory for the solutions of $^3$He in $^4$He. Direct and exchange interactions.
    11.1.1. Three-dimensional case. Spin diffusion measurements.
    11.1.2. Two possible approaches to the fermionic superfluidity in the solutions.
    11.1.3. Three-dimensional Fermi-gas with attraction.
    11.1.4. Three-dimensional Fermi-gas with repulsion.
11.2. Two-dimensional case. $^3$He submonolayers.
    11.2.1. Surface Andreev levels.
    11.2.2. Superfluid thin $^4$He films.
    11.2.3. Spin-susceptibility of $^3$He-submonolayer.
    11.2.4. Possibility of the superfluid transition in the two-dimensional solution.
    11.2.5. Two-dimensional Fermi-gas with attraction.
    11.2.6. Two-dimensional Fermi-gas with repulsion.
11.3. Superfluidity in polarized solutions.
    11.3.1. Three-dimensional polarized solutions.
    11.3.2. Two-dimensional polarized solutions.
11.4. Experimental situation and limitations on the existing theories.
11.5. Two-dimensional monolayers as a bridge between superfluidity and high-$T_C$ superconductivity.

Reference list to Chapter 11.

In this Chapter we will discuss fermionic superfluidity in the Fermi-Bose mixture (in the solution) of $^3$He in $^4$He both in three- and two-dimensional case. We will present the foundations of the classical Bardeen-Baym-Pines theory [11.1] as well as Bashkin-Meyerovich [11.2] results for fermionic s-wave pairing in unpolarized $^3$He-$^4$He mixtures in 3D case, as well as in 2D case for $^3$He submonolayers on Andreev levels [11.3] (which are formed on the free surface of superfluid $^4$He with vacuum) and on grafoil substrates [11.4, 11.17]. We also present the Fermi-gas approach to the p-wave fermionic superfluidity in $^3$He-subsystem based on Kohn-Luttinger [11.5, 11.30, 11.9, 11.7, 11.6, 11.18] mechanism for the pairing of two $^3$He quasiparticles via polarization of the fermionic background.

We will illustrate that the critical temperatures of the p-wave pairing can be strongly increased in a spin-polarized case both for 3D and especially for 2D situation [11.6] and discuss the possible experimental test of the proposed theory.

11.1. Bardeen-Baym-Pines theory for the solutions of $^3$He in $^4$He. Direct and exchange interactions.

One of the most interesting and still experimentally unresolved problems in low-temperature physics is the search for fermionic superfluidity in three-dimensional and particularly in two-dimensional (thin films, submonolayers) [11.6] solutions of $^3$He in $^4$He. In this subsection we will concentrate on new experimental approaches and theoretical results that have been published on this topic. We will stress particularly the role of thin $^3$He films and submonolayers as ideal two-dimensional systems for experimental verification of various theories actual in connection with the problem of high-$T_C$ superconductivity.

Note that a solution of $^3$He in $^4$He is the simplest low-density Fermi-system of $^3$He atoms in an inert superfluid $^4$He condensate, which makes a solution of this kind an ideal object for the development and testing of methods belonging to the realm of Fermi-liquid theory. These methods have been used successfully in describing the normal properties of the solutions (thermodynamic characteristics, transport coefficients) [11.2] and in the prediction of possible superfluidity of the $^3$He subsystem in such solutions [11.1, 11.26, 11.27]. The first classical theory of superfluidity of three-dimensional solutions was proposed by Bardeen, Baym and Pines (BBP) in 1967 [11.1]: they established an elegant analogy between pairing of two $^3$He atoms in a solution via the polarization of the $^4$He background (exchange of virtual phonons) and the electron-phonon interaction in the Bardeen, Cooper and Schrieffer (BCS) theory of superconductivity (Fig. 11.1). In accordance with the ideas of Bardeen, Baym and Pines, the total interaction between two $^3$He particles in a solution consists of two components, direct and exchange:

$$V(r) = V_{dir}(r) + V_{exch}(r). \qquad (11.1.1)$$

The direct interaction includes the contribution of hard-core repulsion at short distances ($V_1$) and of the Van der Waals attraction ($V_2$) at large distances:

$$V_{dir}(r) = V_1(r) + V_2(r). \qquad (11.1.2)$$

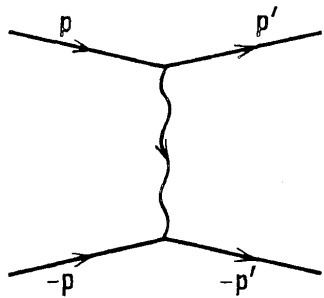

Fig. 11.1. Interaction of two $^3$He atoms via the polarization of the superfluid $^4$He background [11.6].

The exchange interaction $V_{exch}(r)$ represents the interaction of two $^3$He atoms via a local change in the density of $^4$He. This is an analogue of the deformation potential in the BCS theory.

The corresponding expression in the momentum space is:

$$V(q) = V_{dir}(q) + V_{exch}(q), \qquad (11.1.3)$$

where $V_{exch}(q)$ is due to the exchange of a virtual phonon in the three-dimensional case and the exchange of a quantum of third sound in the two-dimensional situation [11.28, 11.29].

At low temperatures and concentrations the subsystem of $^3$He atoms is a low-density Fermi-liquid, i.e. it is effectively a Fermi gas. Therefore, the superfluid transition in this liquid is described by the BCS theory and it depends decisively on the amplitude and the sign of the total interaction $V(q)$ on the Fermi surface. More rigorously, we have $\vec{q} = \vec{p} - \vec{p}'$, where $\vec{p}$ and $\vec{p}'$ are the momenta of the incoming and outgoing particles in the Cooper channel, $|\vec{p}| = |\vec{p}'| = p_F$, and $q^2 = 2p_F^2(1 - \cos\theta)$, $\theta = \widehat{\vec{p}\vec{p}'}$. Thus the only quantity which must be determined when we deal with the Cooper problem is the value of the s-wave harmonic of the potential $V(q)$ on the Fermi surface:

$$V_{l=0} = \int_{-1}^{1} V(q(\cos\theta)) \frac{d\cos\theta}{2}. \qquad (11.1.4)$$

11.1.1. Three-dimensional case. Spin diffusion measurements.

The deformation potential has the following form in the momentum space:

$$V_{exch}(q) = g_q^2 \frac{2\omega_q}{(\varepsilon_p - \varepsilon_{p-q})^2 - \omega_q^2}, \qquad (11.1.5)$$

where $g_q$ is the coupling constant and $\omega_q$ is the frequency of the phonon spectrum of $^4$He. If $|\varepsilon_{p+q} - \varepsilon_p| < \omega_q < \omega_D$ (where $\omega_D$ is Debye frequency), we find that $V_{exch}(q) = -\frac{2g_q^2}{\omega_q} < 0$. In complete analogy with the BCS theory we have $g_q^2 \sim q$, $\omega_q = sq$, where $s$ is the sound velocity in $^4$He, so that the final result is $V_{exch}(q \to 0) = const$. In the case of the solutions this constant is $-(1+\alpha)^2 \frac{m_4 s^2}{n_4} < 0$, where $\alpha \approx 0.28$ is the relative increase in the volume of the solution owing to the replacement of a $^4$He atom with a $^3$He atom; $n_4$ and $m_4$ are the density and mass, respectively, of $^4$He. We should note that in the low-density case (for small concentration of $^3$He in the solution) we have $\omega_D > \varepsilon_F$ and the whole volume of the Fermi sphere (and not only the Debye shell) participates in the superconductive pairing in contrast with the standard BCS theory.

The direct interaction of $^3$He atoms in the momentum space is found from the thermodynamic identity describing the derivative of the chemical potential with respect to the density and has the following form:

$$V_{dir}(q=0) = \frac{\partial \mu_{3\uparrow}}{\partial n_{3\downarrow}} = (1+2\alpha)\frac{m_4 s^2}{n_4} > 0, \qquad (11.1.6)$$

where $\mu_{3\uparrow}$ and $n_{3\downarrow}$ represent, respectively, the chemical potential of $^3$He atoms with "up" spin and the density of $^3$He atoms with "down" spin. The result is:

$$V(q=0) \equiv V_{l=0} = V_{exch}(q=0) + V_{dir}(q=0) = -\alpha^2 \frac{m_4 s^2}{n_4} < 0. \qquad (11.1.7)$$

We can therefore conclude that at very low $^3$He concentrations (when $p_F \to 0$ and, consequently, $q \to 0$) the total interaction is attractive and we can expect the spherically symmetric singlet s-wave pairing which is standard in the BCS theory.

However, spin diffusion experiments show that the situation is far from trivial (see [11.26] and the references therein). In these experiments the dependence of $DT^2$ ($D$ is the spin diffusion coefficient and $T$ is the temperature) of the $^3$He concentration was determined. The experimental curves in [11.26] are strongly non-monotonic and exhibit a maximum at a certain concentration $x_0$ approximately equal to 4% (see Fig. 11.2). They are approximately described by the expression:

$$DT^2 \propto \frac{x^{2/3}}{V_{l=0}^2 - \frac{2}{3}V_{l=0}V_{l=1} + \frac{11}{35}V_{l=1}^2}. \qquad (11.1.8)$$

A theoretical analysis of these experimental curves shows that the absolute value of the s-wave harmonic of the total potential $V_{l=0}$ decreases with an increase in the concentration $x$, than vanishes at $x = x_0$, and at higher concentrations becomes repulsive. On the other hand, for $x \geq x_0$ the p-wave harmonic of the total potential $V_{l=1}$ is significant and attractive (although smaller than $V_{l=0}$ at $x = 0$). These circumstances lead to the two possible approaches to the fermionic superfluidity in the solution.

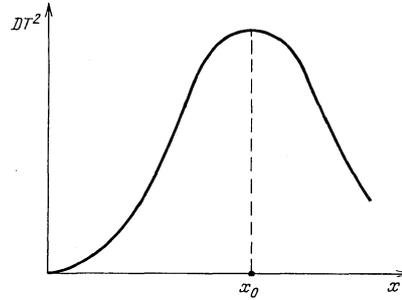

Fig. 11.2. Approximate experimental dependence of the product $DT^2$, representing the spin diffusion in a solution, on the concentration $x$ of $^3$He [11.6, 11.26].

11.1.2. Two possible approaches to the fermionic superfluidity in the solutions.

In the first approach it is assumed that the total interaction of two $^3$He atoms described by $V(q)$ exhibits significant momentum dependence and, moreover, its sign is reversed at the values of the vector $q$ of the order of the Fermi momentum when the concentration is $x_0$, i.e. when it is $p_F(x_0)$. This hypothesis leads to the model potential of the BBP theory [11.1]:

$$V(q) = V(q=0)\cos\frac{q}{k_s}, \quad k_s \sim p_F(x_0). \qquad (11.1.9)$$

The BBP model potential was improved in 1989 by van de Haar, Frossati and Bedell [11.26]. They introduced the concentration dependence of the amplitude of the potential $V(q = 0)$:

$$V(q=0) = -\frac{m_4 s^2}{n_4}\alpha^2(1+\gamma\frac{x}{x_{max}}), \qquad (11.1.10)$$

where $x_{max}$ is the solubility limit of $^4$He at a given pressure $P$ and $\gamma(P)$ is a fitting parameter. In both theories [11.1] and [11.26] the s-wave harmonic of the total interaction is maximal and attractive at low concentrations and then it begins to decrease in absolute value, changing sign to become a repulsive one at concentrations corresponding to $p_F \sim k_s$. At higher concentrations the p-wave harmonic of $V(q)$ becomes attractive. Therefore, van de Haar, Frossati and Bedell predict singlet s-wave pairing in a solution at low concentrations of $^3$He and triplet p-wave pairing at high concentrations. It should be pointed out that the two fitting parameters $k_s = k_s(p)$ and $\gamma(P)$, extracted from the experiments on spin diffusion and magnetostriction, are used in the improved model potential of van de Haar, Frossati and Bedell.

The second approach, adopted by us and by others [11.6, 11.2, 11.27], does not rely on any model potential. In this approach the only microscopic parameter of the system is the s-wave scattering length $a = \dfrac{m}{4\pi} V_{l=0}$, which is dependent on the pressure and concentration. It is assumed that its sign is reversed at a concentration corresponding to the maximum of the $DT^2$ curve (see Fig. 11.3).

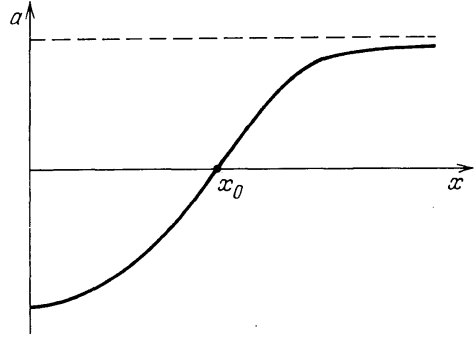

Fig. 11.3. Qualitative dependence of the scattering length in a solution from the concentration of $^3$He. At $x = 100\%$ the value of $a(x)$ tends to the scattering length of pure $^3$He and is approximately equal to $2/p_{F0}$ at zero pressure [11.6] (here $p_{F0}$ is the Fermi momentum of pure $^3$He).

It should be pointed out that the higher harmonics $(V_{l=1}, V_{l=2},...)$ appear in the second order of the perturbation theory but not because of the momentum dependence of the total interaction $V(q)$: they originate from the scattering length $a$ because of the effective interaction of two $^3$He particles via the fermionic background of their own $^3$He subsystem.

The relationship between these two approaches is approximately the following. Let us assume, for the sake of simplicity, that the direct interaction of two $^3$He particles in a solution is described by:

$$V_{dir}(r) = \begin{cases} V_1; & r < r_1, \\ -V_2; & r_1 < r < r_2, \end{cases} \qquad (11.1.11)$$

where the first term is responsible for the hard-core repulsion at short distances and the second term is due to the Van der Waals attraction at long distances (see Fig. 11.4 and Fig. 6.1).

At low $^3$He concentrations in a solution, i.e. in the case when $p_F r_1 \ll p_F r_2 \ll 1$, the s-wave harmonic of the direct interaction is $V_{dir}^{l=0} = V_1^{l=0} - V_2^{l=0}$. Then, if

$$V_1^{l=0} - V_2^{l=0} - V_{exch}^{l=0} < 0, \qquad (11.1.12)$$

but

$$V_1^{l=0} - V_{exch}^{l=0} > 0, \qquad (11.1.13)$$

we have the low-density Fermi gas with the gas parameter $p_F r_1 \ll 1$ and with a scattering length which changes its sign at $p_F \ll 1/r_2$. Naturally, this approach ignores the p-wave harmonic of the Van der Waals interaction, which need not be small in the transition region $p_F r_2 \sim 1$. It should be pointed out that at high concentrations when $p_F r_2 \gg 1$ we find that $V_2^{l=1}$ is small and of the same order as $V_2^{l=0}$. In this Chapter the second (Fermi-gas) approach to the problem of superfluidity in the solutions will be mainly used.

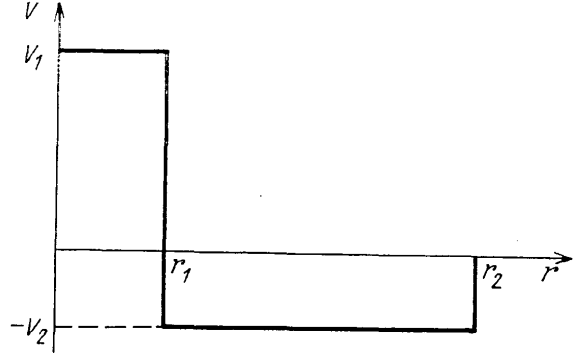

Fig. 11.4. Model representation of the direct interaction of two particles in a solution as a function of the distance *r* between them [11.6].

11.1.3. Three-dimensional Fermi-gas with attraction.

The expression for the temperature of the superfluid transition in a Fermi gas with attraction was first obtained by Gor'kov and Melik-Barkhudarov in 1961 [11.19], soon after the creation of the BCS-theory. Bashkin and Meyerovich [11.2] used this expression to describe the superfluidity of the solutions at very low concentrations of $^3$He. For the concentrations in the range $x < x_0$ and an attractive s-wave scattering length $a < 0$ the expression for the s-wave critical temperature reads:

$$T_{C0} = 0.28\varepsilon_{F0} x^{2/3} \exp\left(-\frac{\pi}{2|a|p_{F0}x^{1/3}}\right), \qquad (11.1.14)$$

where $\varepsilon_{F0}$ and $p_{F0}$ are the Fermi energy and momentum of pure $^3$He. It is worth noting that the preexponential factor in this expression is proportional to $\varepsilon_F$ and not $\omega_D$, as in the case of the phonon model.

According to the estimates of Østgaard and Bashkin [11.36], the maximum value of $T_{C0}$ is $T_{C0}(1\%) \sim 10^{-4}$ K. Frossati and his colleagues [11.26] proposed a lower critical temperature with maximal $T_{C0} = T_{C0}(2\%) \sim (4 \cdot 10^{-6} \div 10^{-5})$ K. In their estimates they obtained the larger value of $T_{C0}$ of the order of $10^{-5}$ K by extracting the fitting parameters from the magnetostriction experiments, and $4 \cdot 10^{-6}$ K from the spin diffusion experiments. At a given concentration $x$ the gas parameter of the theory is $ap_{F0}x^{1/3}$ and it depends weakly on the pressure.

11.1.4. Three-dimensional Fermi gas with repulsion.

At higher concentrations ($x > x_0$) the scattering length changes its sign $a > 0$, and s-wave pairing becomes impossible. Nevertheless, even in this case the subsystem of $^3$He atoms in a solution may become a superfluid, but this is now due to an instability with respect to triplet p-wave pairing. As we already discussed in Chapter 9, the mechanism for the realization of p-wave pairing was first considered by Fay and Layzer [11.18] and Kagan, Chubukov [11.6] following pioneering ideas of Kohn and Luttinger [11.5]. It is related to the presence of Kohn's anomaly (of the Friedel oscillations [11.41]) in the effective interaction of Fermi particles via the polarization of the Fermionic background. As a result the purely repulsive short-range potential between two particles in vacuum gives rise to an effective interaction in matter with the competition between attraction and repulsion in it. A rigorous calculation [11.18, 11.6] shows that for all the harmonics of the effective potential (except the s-wave harmonic) the attraction wins in this competition and the p-wave harmonic is the most attractive. Consequently, a three-dimensional Fermi gas with repulsion is unstable with respect to the superfluid transition with the triplet p-wave pairing below the critical temperature:

$$T_{C1} \sim \varepsilon_{F0} x^{2/3} \exp\left\{-\frac{5\pi^2}{4(2\ln 2 - 1)a^2 p_{F0}^2 x^{2/3}}\right\}, \qquad (11.1.15)$$

where maximal $T_{C1}$ corresponds to the pressures $P = 10$ bar, when we have the maximal solubility of $^3$He $x_{max} = 9.5\%$. In this case it equals to $(10^{-10}\text{-}10^{-9})$ K. The triplet pairing temperature of this order of magnitude was predicted also by Bardeen, Baym and Pines [11.1]. Frossati and others [11.26] give a more optimistic estimate for the triplet pairing case. At the maximal concentration $x_{max} = 9.5\%$ the value of $T_{C1}$ lies between $10^{-6}$ and $10^{-4}$ K. The lower temperature of the p-wave pairing ($10^{-6}$ K) is obtained in [11.26] when the fitting parameters are extracted from transport experiments and the higher temperature ($10^{-4}$ K) follows from magnetostriction experiments. Note that at zero pressure the maximal solubility of $^3$He in the solution is only 6.4% [11.1, 11.2].

11.2. Two-dimensional case. $^3$He-submonolayers.

A solution of $^3$He in $^4$He is also very interesting because it can be made purely two-dimensional. In superconducting electron systems a film is regarded as a two-dimensional if its thickness $L$ is much less than the coherence length $\zeta_0 \sim 1000$ Å. In $^3$He films on grafoil (exfoliated graphite) and in monolayers and submonolayers the radius of the localization of $^3$He atoms in the third dimension (which is the film thickness) is indeed of the order of the distance between the atoms $d$. Therefore, by analogy with inversed layers in heterostructures, we are also dealing here with a purely two-dimensional system, and, moreover, our system is free of impurities. In this sense a two-dimensional solution of $^3$He in $^4$He can be regarded as a bridge between superfluidity and superconductivity, particularly high-temperature superconductivity. In fact, the majority of modern theories of high-$T_C$ superconductivity rely on the two-dimensional or quasi-two-dimensional behavior (see Chapters 11, 12, 13 and 16) for the unusual normal properties (resistivity, susceptibility, small $Z$-factor) of these materials, as well as to account for the high temperature of their superconducting transition. Two-dimensional helium films and particularly monolayers and submonolayers with a low two-dimensional $^3$He density are ideal objects for experimental verification of the different fashionable theories of high-$T_C$ superconductivity, such as the theory of a marginal Fermi-liquid (see Chapter 16) proposed by Varma et al [11.31] or a somewhat similar theory of the Luttinger liquid proposed by Anderson [11.32]. These topics will be discussed again at the end of this subsection. At this stage we will provide a brief review of the history of the experimental discovery and theoretical prediction of the existence of a two-dimensional solution.

11.2.1. Surface Andreev Levels.

The first experiments were carried out by Esel'son and Bereznyak [11.33] and by Atkins and Narahara [11.34]. These experiments revealed a nontrivial temperature dependence of the surface tension (in fact, the surface free energy) of a weak solution of $^3$He in $^4$He. The experiments were interpreted by Andreev [11.3] who postulated the existence of surface $^3$He-impurity levels on the free surface of superfluid $^4$He with vacuum (which are similar to some extent to Tamm's surface layers in metals). This idea was subsequently confirmed by detailed experiments of Zinov'eva and Boldarev [11.35] and of Edwards et al [11.37] as well as by the variational calculations (cf. the review of Edwards and Saam [11.38] and the literature cited there). The correct interpretation of the experimental results yields the following parameters representing the surface state: $\varepsilon = -\Delta - \varepsilon_0 + \dfrac{p_\parallel^2}{2m^*}$, where $\Delta = 2.8$ K is the binding energy of a $^3$He quasiparticle in the bulk (Andreev [11.3], Bashkin and Meyerovich [11.2]), $\varepsilon_0 = 2.2$ K is the difference between the binding energies of a $^3$He quasiparticle in the bulk and on the surface; $m^* = 1.5 m_3$ is the hydrodynamic effective mass governing the motion of $^3$He quasiparticles along the surface. It should be pointed out that, according to the variational calculations of Lekner [11.39] and Saam [11.40], the appearance of the Andreev levels is due to a combination of the effects associated with the Van der Waals interaction between $^3$He and the $^4$He density profile (which varies when we approach the free surface) and with the difference between the energies of the zero-point motion of $^3$He and $^4$He. Such effects lead to the localization of $^3$He atoms near the free surface. In the same time $^3$He atoms can move freely along the surface of $^4$He, which is almost equipotential because the hydrodynamic condition $\mu_4 = const$ (see Chapter 1) is satisfied on this surface. The wave function of the Andreev state is $\Psi = \Psi(z)\exp(i\vec{p}_\parallel \vec{r}_\parallel)$ with $\Psi^-(z) \sim \exp(-z/d)$, where $d$ is the radius of localization along the normal to the surface.

### 11.2.2. Superfluid thin $^4$He-films.

The first experiments on thin $^4$He films (film thickness is less than 25 Å) of the same kind as the experiments of Zinov'eva and Boldarev [11.35] and of Edwards et al [11.37], were carried out by Gasparini, Bhattacharyya, and Di Pirro [11.42]. Gasparini and others determined the contribution of the surface states of $^3$He to the specific heat of the thin films. They also proposed the first theoretical interpretation of the results [11.43]. Subsequently several experimental papers were published by Hallock et al [11.4, 11.44, 11.45], who measured the magnetization and the spin-lattice relaxation time of $^3$He submonolayers formed on the surface of thin $^4$He films.

The theoretical interpretation of the experiments of Hallock et al proposed by Dalfovo and Stringari [11.46], Pavloff and Treiner [11.47], Krotscheck, Saarela and Epstein [11.48] require the assumption that not one but two Andreev levels exist on the surface of a thin $^4$He film. The energy of the first Andreev layer, $E_1 = -\Delta - \varepsilon_1 + \frac{p_\parallel^2}{2m_1}$, is practically identical with the energy of the Andreev level ($\varepsilon_1 \cong \varepsilon_0$) on a bulk surface, differing only in respect of the effective mass $m_1 \approx 1.35 m_3$. The energy of the second Andreev level is still lower than the energy of $^3$He in the bulk and is given by the expression $E_2 = -\Delta - \varepsilon_2 + \frac{p_\parallel^2}{2m_2}$, where in the limit of zero concentration of $^3$He and not too thin films we have $m_2 \approx 1.6 m_3$ and $\varepsilon_2 \approx 0.4$ K; consequently, $\varepsilon_2 - \varepsilon_1 = 1.8$ K.

The wavefunction of the first Andreev level is localized mainly near the free surface and has a significant tail (~3 Å) above the surface. At the same time the wavefunction of the second Andreev level penetrates partly into the film. Two Andreev levels (instead of one) appear according to [11.46-11.48] because of the competition between the size effect (vanishing of the $\Psi$-function of $^3$He near the substrate and consequent increase of the kinetic energy $E_{kin} \sim (\nabla_z \Psi)^2 \sim 1/L^2$ of $^3$He) and the Van der Waals attraction by the substrate (which is proportional to $1/L^3$ and tends to reduce the energy). In the case of thin and moderately thick films the Van der Waals attraction is stronger than the size-effect repulsion and, therefore, the energy of the second Andreev level is still lower than the energy of $^3$He in the bulk.

In the case of very thick films the Van der Waals attraction of the substrate (proportional to $1/L^3$) may become unimportant compared with the kinetic energy, and the energy of the second Andreev level may prove to be higher than the energy of $^3$He in the bulk. In this case the second level evidently vanishes by merging with the bulk levels. At a fixed film thickness, the Van der Waals attraction of the substrate depends on whether the substrate is "strong" or "weak". On a weak substrate (Cs, Rb, K, Na, Li, Mg, H$_2$) it is found that $^4$He is in the liquid phase. On a strong substrate (Ag, Au, Cu, Al) one or two solid $^4$He layers are formed on it, and $^4$He is in a liquid phase only starting from the third and following layers. The presence of one or two solid layers reduces the Van der Waals attraction of the substrate and increases the kinetic energy, leading to a possible disappearance of the second Andreev level at lower thicknesses of the film.

We would like to emphasize that the topic of surface levels in the films is not fully understood yet. There is an alternative point of view according to which the second Andreev level can exist not only in thin films, but also in the bulk.

It is thus clear that in the case of not very thin and not very thick films there are definitely two Andreev levels whose energies differ by $\varepsilon_2 - \varepsilon_1 = 1.8$ K. Their existence is manifested in Hallock experiments by the presence of a step in the dependence of the magnetization on the surface density of $^3$He. This step appears when the density of $^3$He is equal to 0.85 of a monolayer. At lower densities the second Andreev level is not important, and we are dealing with a purely two-dimensional one-level system whose spectrum is $E = -\Delta - \varepsilon_1 + \frac{p_\parallel^2}{2m_1}$ and the wave-function is $\Psi = \Psi(z)\exp(i\vec{p}_\parallel \vec{r}_\parallel)$.

### 11.2.3. Spin susceptibility of $^3$He-submonolayers.

Another important result reported by Hallock et al is an analysis of the temperature dependence of the susceptibility. At low temperatures ($T \ll T_F$) the susceptibility depends weakly on temperature, and for surface densities from 0.03 to 0.3 of a monolayer it is well described by an expression for a two-dimensional Fermi gas with weak repulsive interaction (with coupling constant $f_0$) between the particles [11.61, 11.62]:

$$\chi = \chi_0 \frac{1 + \frac{1}{2}F_1^s}{1 + F_0^a} \sim \chi_0(1 + f_0), \qquad (11.2.1)$$

where $\chi_0$ is susceptibility of the non-interacting 2D Fermi-gas, $F_1^s \sim f_0^2$ and $F_0^a \sim f_0$ are two-dimensional harmonics of the Landau quasiparticle interaction function, $f_0 \sim \frac{1}{2\ln 1/p_F r_0}$ is a two-dimensional coupling constant (introduced in Chapter 11) and $r_0$ is the range of the potential.

At the surface densities from 0.005 to 0.03 we have $\chi < \chi_0$, which supports the attractive sign of the coupling constant (this is also true for the gas-parameter $a(x)p_{F0}x^{1/3}$ in the case of three-dimensional solutions). Note that the exact surface densities at which the 2D coupling constant changes its sign can be determined from the measurements at lower temperatures $T < T_F$, since $T_F \sim T_{F0}x$ is small and the transition from the Fermi gas behavior of the susceptibility to the Curie law occurs very early.

Concluding this subsection we must mention that there is also another purely two-dimensional system, namely $^3$He on the surface of grafoil which has very similar properties at a low surface density of $^3$He (cf. experiments carried out by Saunders group [11.17, 11.59]). The rest of our discussion of fermionic superfluidity in purely two-dimensional low-density systems can be also applied to 3He on grafoil, subject only to small modifications.

### 11.2.4. Possibility of the superfluid transition in the two-dimensional solutions.

We shall now consider the possibility of the superfluid transition in a two-dimensional $^3$He submonolayers on the surface of $^4$He.

By analogy with the tree-dimensional case, the total interaction between two $^3$He particles on the surface is given by the expression:

$$V(\vec{r}, z) = V_{dir}(\vec{r}, z) + V_{exch}(\vec{r}, z), \qquad (11.2.2)$$

where the exchange interaction $V_{exch}(\vec{r}, z)$ is governed by the sum of the residual parts (which are not used to form Andreev level) of the deformation potential of the interaction between two $^3$He particles via the polarization of $^4$He, and of the Van der Waals attraction of the substrate.

These residual parts of the exchange interaction are related primarily with the interaction of $^3$He particles with the curved surface of superfluid $^4$He in the field of surface waves of a third sound. The spectrum of third sound waves is of the form $\omega^2 = \frac{\alpha}{\rho}(\kappa^2 + q^2)q \cdot \text{th}(qL)$ [11.49, 11.50], where the first term in the parentheses describes the contribution of the Van der Waals potential of the substrate and the second one represents the local surface change in the density of $^4$He. In the case of thin films the contribution of the first term predominates, i.e. the dynamic part of the Van der Waals potential of the substrate is more important than "surface phonons" (which are called riplons). Consequently, a reduction in the film thickness changes the spectrum from the purely riplon type $\omega^2 = \frac{\alpha}{\rho}q^3$, where $\alpha$ is the surface tension [11.51] to an acoustic spectrum with a linear dispersion law [11.49] $\omega^2 = \frac{\alpha\kappa^2}{\rho}q^2$, where $\kappa$ is the capillary constant of the Van der Waals potential and $L$ is the film thickness $qL \ll 1$.

In the two-dimensional problem it is important, as always, to determine the two-dimensional projection of the three-dimensional potential $V(\vec{r}, z)$. In close analogy with the two-dimensional projection of the Coulomb interaction (briefly considered in Chapter 12 in the momentum space), in the real space it is given by the expression:

$$V(\vec{r}_1 - \vec{r}_2) = \iint V(\vec{r}_1 - \vec{r}_2, z_1 - z_2)|\Psi(z_1)|^2|\Psi(z_2)|^2 dz_1 dz_2, \qquad (11.2.3)$$

where $\Psi(z)$ is the wave function of Andreev level. The two-dimensional projection of the total interaction can be represented in the form:

$$V(r) = V_{dir}(r) + V_{exch}(r), \qquad (11.2.4)$$

and correspondingly in the momentum space

$$V(q_\|) = V_{dir}(q_\|) + V_{exch}(q_\|). \qquad (11.2.5)$$

In (11.2.5) $V_{exch}(q_\|)$ is due to the exchange of a virtual quantum of the third sound. Note that in the case of thin films $qL \ll 1$: $V_{exch}(q_\| = 0)$ is given by $V_{exch}(q_\| = 0) = -m_4 c_{III}^2$ [11.28, 11.29], where $c_{III}$ is the velocity of the third sound. It is given by:

$$c_{III}^2 = \frac{3V_{sub}h_4}{m_4 n_4 (d + h_4)^4} \qquad (11.2.6)$$

In (11.2.6) $V_{sub}$ is the amplitude of the Van der Waals potential of the substrate, $d$ and $h_4$ are the thicknesses of the solid ($d$) and the superfluid ($h_4$) layers, so $L = d + h_4$ is a total film thickness. (We recall that in the case of well-wetted substrate such as Au, Ag, Cu, etc, the first $^4$He layer on the substrate solidifies).

It should be pointed out that, as in the three-dimensional problem, the limiting frequency of surface waves $\omega_D \sim m_4 c_{III}^2$ is much higher than the Fermi energy $\varepsilon_F$. Therefore, we are again dealing with an antiadiabatic situation, when the whole volume of the two-dimensional Fermi sphere (and not only its Debye shell) is important in the problem of the possibility of a superfluid transition.

We shall now consider the direct part of the total interaction. By analogy with the three-dimensional case, we have:

$$V_{dir}(q_\| = 0) = V_1(q_\| = 0) + V_2(q_\| = 0), \qquad (11.2.7)$$

where $V_1$ is determined by the hard-core repulsion between two $^3$He particles at short distances, whereas $V_2$ is due to the attractive interaction of two $^3$He particles at large distances (see Fig. 11.4).

As pointed out at the end of the preceding subsection, the experiments of Hallock et al [11.4, 11.44, 11.45] on the dependence of the magnetization of a submonolayer on the surface density of $^3$He demonstrate that the total interaction of two $^3$He particles on the surface of a thin film is attractive when the ³He concentration is small ($x < 3\%$), and repulsive when the concentration is higher. Therefore, the direct part of the total interaction can be represented again in a model form shown in Fig. 11.4, i.e.:

$$V_{dir}(r) = \begin{cases} V_1, & r < r_1 \\ -V_2, & r_1 < r < r_2 \end{cases}, \qquad (11.2.8)$$

where $1/r_2 \sim p_F(x=3\%)$. If the submonolayer density is such that $p_F^2 < \dfrac{1}{r_1^2}$ ($p_F r_1 < 1$), we can try to use the Fermi-gas approach again. It should be pointed out, that the experiments of Hallock et al demonstrate, that at ³He densities less than 0.3 of a monolayer ($x < 30\%$) we are indeed dealing with a weakly interacting low-density two-dimensional Fermi gas.

11.2.5. Two-dimensional Fermi gas with attraction.

A special feature of the two-dimensional case is that for purely attractive potential (without hard-core repulsion part) even if the attraction is infinitely weak, we are dealing with the coexistence of two phenomena [11.52 – 11.54]: pairing of two particles in vacuum in the coordinate space and the Cooper pairing of two particles in substance in the momentum space in the presence of a filled Fermi sphere. In the case of a purely attractive potential the energy of a bound state in vacuum in weak-coupling case reads:

$$E_b = -\frac{1}{mr_0^2}\exp\left(-\frac{4\pi}{m|U_0|}\right), \qquad (11.2.9)$$

where $U_0$ is the s-wave harmonic of the potential and $r_0$ is the range of the potential. In the weak-coupling case $|f_0| = \dfrac{m|U_0|}{4\pi} \ll 1$ for the 2D gas-parameter $f_0$. The s-wave critical temperature according to Miyake is $T_{C0} = \sqrt{2\varepsilon_F |E_b|}$ for $\varepsilon_F \gg |E_b|$ (see Chapter 14).

The situation is more complicated in the case of the potential with a repulsive core and an attractive tail (see Fig. 11.4). Here as it was shown in [11.55] in the case of strong hard-core repulsion $V_1 \gg \left\{|V_2|; \dfrac{1}{mr_0^2}\right\}$ we have a threshold for s-wave two-particle pairing. The first solution for the bound state for $k = \sqrt{m|E_b|} \to 0$ appears when $\kappa r_1 \geq 0.4$ and $\kappa r_2 \geq 1.6$, where $\kappa = \sqrt{m(|V_2|-|E_b|)}$. It means that an attractive tail must be at least four times more extended than a hard-core region ($r_2 \geq 4r_1$). The coordinate part of the Ψ-function in this case monotonously increases in the interval between $r_1$ and $r_2$ (in close analogy with a solution for the bound-state of the extended s-wave pairing in the low-density 2D t-J model with released constraint considered in Chapter 13 or with a lattice gas with Van der Waals potential considered in Chapter 5). Thus we can say, that $r_2$ is the mean distance between the two particles. The threshold condition can be represented as $m|V_{2C}|r_2^2 \approx 2.56$. Note that $mr_2^2$ corresponds to $1/2t = md^2$ in a lattice model. It means that a threshold in a continuous model is larger than in a lattice model (see Chapter 5). Note that the coordinate part of the two-particle Ψ-function does not depend upon the statistics of two particles – it is the same for two fermions with spins "up" and "down" in singlet ($S_{tot} = 0$) state considered in Chapter 13 and for two spinless and structureless bosons considered in Chapter 5. Hence the situation in continuous Van der Waals model remains qualitatively the same as in the lattice model (see Fig. 6.1 and [11.55]).

The equation for the energy spectrum reads (see [11.55]):

$$\frac{\kappa(Y_0(\kappa r_1)J_1(\kappa r_2)-J_0(\kappa r_1)Y_1(\kappa r_2))}{Y_0(\kappa r_1)J_0(\kappa r_2)-J_0(\kappa r_1)Y_0(\kappa r_2)} = \frac{kK_1(kr_2)}{K_0(kr_2)}, \qquad (11.2.10)$$

where $k = \sqrt{m|E_b|}$ and $\kappa = \sqrt{m(|V_2|-|E_b|)}$, $J_0$, $Y_0$ and $J_1$, $Y_1$ are the Bessel functions of zeroth and first orders $K_0$ and $K_1$ are the Macdonald functions of zeroth and first order.

For shallow bound state $|E_b| << \frac{1}{mr_2^2}$ close to the threshold $\frac{|V_2|-|V_{2C}|}{|V_2|} << 1$ we have (see [11.56]):

$$|E_b| \sim \frac{1}{mr_2^2} \exp\left\{-\frac{1}{\lambda}\right\}, \qquad (11.2.11)$$

where $\lambda = \frac{|V_2|-|V_{2C}|}{\pi|V_2|}$. When the amplitude of the Van der Waals attraction strongly exceeds $|V_{2C}| \approx \frac{2.56}{mr_2^2}$, the binding energy $|E_b| \leq \frac{1}{mr_2^2}$. Note that the famous Miyake formula is still valid for the potential with the hard-core repulsion and the Van der Waals attractive tail in the weak coupling case $|E_b| << \varepsilon_F$. In this case $|E_b| << T_C << \varepsilon_F$. In the opposite strong-coupling (or very diluted) case when $|E_b| >> \varepsilon_F$, as we discussed in Chapters 5 and 14, we have two characteristic temperatures: the bound pairs are formed at the crossover Saha temperature [11.57] $T_* \sim \frac{|E_b|}{\ln\frac{|E_b|}{\varepsilon_F}}$, while the pairs are Bose-condensed at Fisher-Holenberg s-wave critical temperature [11.58]

$$T_{C0} = \frac{\varepsilon_F}{2\ln\ln\frac{|E_b|}{\varepsilon_F}}.$$

We note that even in the strong-coupling case $T_{C0} \leq \varepsilon_F$, where $\varepsilon_F = \varepsilon_{F0}x$ and $x$ is the 2D $^3$He concentration. We shall conclude this subsection with a rough estimate for the s-wave temperature of the two-dimensional superfluid transition. According to Bashkin, Kurihara and Miyake we can expect $T_{C0}$ of the order of 1mK or less when the surface density of $^3$He is of the order of 0.01 of a monolayer.

11.2.6. Two-dimensional Fermi-gas with repulsion.

When the $^3$He surface density exceeds 0.03 of monolayer, the total interaction between $^3$He particles changes sign and the s-wave pairing becomes impossible. However even in this case $^3$He subsystem will become unstable towards triplet p-wave pairing [11.5] below the critical temperature $T_{C1} \sim \varepsilon_F \exp\left\{-\frac{1}{6.1f_0^3}\right\}$ in the third order in $f_0$ for the effective interaction $U_{\text{eff}}(q)$ (see Chapter 9), where $\varepsilon_F = \varepsilon_{F0}x$ and $f_0 = \frac{1}{\ln\frac{1}{p_{F0}^2 x r_0^2}}$ is 2D coupling constant, $r_0 \sim r_1$ is the range of the hard-core part of the potential [11.61, 11.62]. An estimate of the superfluid transition temperature obtained by Chubukov [11.61] is $T_{C0} \sim 10^{-4}$ K for densities $\sim 0.3$ of a monolayer corresponding to the limit of the validity of the Fermi-gas description.

It should be pointed out that the allowance for nonquadratic corrections to the spectrum of $^3$He quasiparticles $\varepsilon(p) = \frac{p^2}{2m}(1-\gamma\frac{p^2}{p_C^2})$, results in the appearance of the p-wave pairing already in a second order of the perturbation theory for effective interaction. This yields $T_{C1} \sim \varepsilon_F \exp\left\{-\dfrac{1}{\gamma f_0^2}\right\}$ as demonstrated by Baranov, Kagan [11.25]. However, the superfluid transition temperature now depends exponentially on the small constant $\gamma$ representing the quartic corrections to the quadratic spectrum of quasiparticles and, therefore, $T_{C1}$ is very small.

### 11.3. Superfluidity in polarized solutions.

We shall now consider briefly the situation in strongly (spin) polarized solutions.

It is well known that the singlet s-wave pairing in a strongly polarized solution is suppressed by a paramagnetic effect. This means that in magnetic fields larger than the paramagnetic limit $H > \dfrac{T_{C0}}{\mu_B}$ (see Chapter 12), where $\mu_B$ is the nuclear Bohr magneton for $^3$He, the singlet s-wave superfluid state is destroyed. The influence of an external magnetic field (or of the spin polarization) on the triplet p-wave pairing temperature is less trivial.

#### 11.3.1. Three-dimensional polarized solutions.

Chubukov and Kagan [11.9] showed that the p-wave pairing temperature of a three-dimensional polarized gas with repulsion depends strongly and nomonotonically on the degree of polarization $\alpha$ (see Chapter 10 and Fig. 10.2): it rises strongly at low and intermediate polarization (for $S_{tot} = S_{tot}^2 = 1$), passes through a maximum when the polarization is 48%, and falls on further increase of polarization. At a pressure of 10 bar the maximal possible concentration of $^3$He is $x = 9.5\%$ and the temperature in the maximum corresponds to $10^{-6}$–$10^{-5}$K [11.6], which is much higher than $T_{C1} \sim 10^{-10} \div 10^{-9}$ K corresponding in [11.6] to the case when $\alpha = 0$. An account of preexponential factors for $T_{C1}$ further increases the critical temperature especially in field [11.24]. A qualitatively similar dependence of $T_{C1}$ on $\alpha$ with a maximum at $\alpha = 32\%$ was also predicted by van de Haar, Frossati and Bedell [11.26]. The temperature at the maximum predicted by these authors for the same values of pressure and concentration is somewhat higher and amounts to $10^{-5}$–$10^{-4}$ K.

The hope for experimental creation of strongly polarized solutions is based primarily on the elegant idea of Castraing and Nozieres [11.21]. In their classical paper they proposed to create a strong polarization in a liquid solution by fast melting of a solid solution. The idea is that a solid solution (and pure crystalline $^3$He) does not have a kinetic energy of the degeneracy of $^3$He atoms associated with the Pauli principle. Therefore, the application of a magnetic field of the order of the Curie temperature:

$\mu_B H \sim T_C \sim T \sim 1\,\text{mK}$ and $H \sim 1\,\text{T}$         (11.2.12)

leads to an almost 100% polarization of the solid solution. It should be pointed out that a significant brute force polarization in a liquid solution can be achieved only by applying very high external magnetic fields $\mu_B H \sim \varepsilon_F \sim 0.1\,\text{K}$ and $H \sim 100$ T. These fields is difficult to reach experimentally.

In the same time fast melting of a strongly polarized ($\alpha \sim 90\%$) solid solution should, according to the estimates of Castraing, Nozieres [11.21], produce a liquid solution with $\alpha \sim 30\%$ which is close to the maximum of $T_C$ in the theory of van de Haar, Frossati and Bedell. Naturally, this polarization is of a nonequilibrium nature, but its lifetime is very long ($\sim 30$ min) because of the long relaxation time in the liquid phase.

Another very important idea for the increase of the critical temperature is the suggestion of Meyerovich et al [11.22, 11.23] according to which the maximal solubility of a strongly polarized solution may be 3-4 times higher than the maximal solubility in the absence of polarization ($x_{\max}(\alpha) \sim 30\%$ instead of $x_{\max}(\alpha = 0) \sim 9.5\%$). A combination of the ideas of Castraing, Nozieres and Meyerovich may produce an even greater increase (to $10^{-4}$–$10^{-3}$ K) in the superfluid transition temperature of a strongly polarized solution.

   11.3.2. Two-dimensional polarized solutions.

   The situation in two-dimensional polarized submonolayers at $^3$He densities from 0.03 to 0.3 of a monolayer is even more favorable from the point of view of the superfluid transition temperature. As it was shown in Chapter 12, the competition between a strong 2D Kohn's anomaly and a reduction of the density of states of the "down" spins (antiparallel to the field) again gives rise to a nonmonotonic dependence of triplet p-wave temperature $T_{C1}$ on the degree of polarization $\alpha$, with a very strong maximum at $\alpha = 60\%$ (see Fig. 10.3 and [11.6, 11.9, 11.10]). It should be pointed out that the maximum is very broad and extends from 10% to 90% of the polarization.
   Estimates in [11.6] indicate that the critical temperature is fairly high now in experimentally achievable fields $H \sim 15$ T. In fact, for the two-dimensional solution with surface density $n_3 \sim 0.05$ of a monolayer and the Fermi-energy $\varepsilon_F \sim 0.13$ K, the application of such a magnetic field produces a polarization degree $\alpha = \dfrac{\mu_B H}{\varepsilon_F} \sim 10\%$. In this case the triplet p-wave pairing temperature (for $S_{tot} = S_{tot}^z = 1$) can reach a value $T_{C1}^{\uparrow\uparrow} \sim 1$ mK, which is quite nice and promising.

   11.4. Experimental situation and limitations on the existing theories.

   In the introduction to the Section 11.1 we pointed out that the search for fermionic superfluidity of $^3$He in three-dimensional and two-dimensional solutions has not resulted in experimental success yet. The published experimental results demonstrating the absence of superfluidity at certain pressures and concentrations impose some limits on the various theoretical estimates of the superfluid transition temperature. They are forcing both theorists and experimentalists to concentrate on those ranges of the parameters where the experiments have not been carried out yet. The review of Østgaard and Bashkin [11.36] contains the experimental results obtained by the groups of Pobell [11.14] and Ogawa [11.20] (we must also mention here more recent experimental results of Saunders group for the 2D case [11.13, 11.63]). They demonstrate the absence of the superfluid transition in the three-dimensional solutions right down to 0.2 mK for $^3$He concentrations of 1%, 5% and 6.4%. They show that the temperature both of the singlet s-wave pairing and the temperature of the triplet p-wave pairing (we recall that the s-wave pairing is suppressed for the $^3$He concentrations exceeding 4% in 3D) most probably lie below 0.2 mK.
   The estimates of Østgaard and Bashkin concerning s-wave pairing in 3D case show that the most promising way is to seek the singlet superfluidity at $^3$He concentrations amounting to ~ 0.5–1%. Van de Haar, Frossati and Bedell [11.26] assume that the optimal concentration lies in the interval 1.5–2.5%. The corresponding temperature $T_{C0}$ is of the order of 0.1 mK for the results of the both groups. According to the estimates of Frossati, Bedell, Meyerovich and Kagan, the triplet superfluidity is most likely to occur at the maximal possible concentration of $^3$He $x_{max} \approx 9.5\%$ which corresponds to the pressure 10 bar, under strong polarization conditions (in strong effective magnetic fields). As pointed out above, when the polarization is very strong, it may be possible to reach $^3$He concentrations even exceeding 9.5%. The most realistic estimates once again (similar to the estimates for an s-wave temperature $T_{C0}$) predict a triplet pairing temperature $T_{C1}$ only of the order of 0.1 mK or lower. Therefore, we obviously can expect that both the singlet and the polarization-enhanced triplet pairing temperatures cannot exceed the value of the order of 0.1 mK.

It seems that the situation in the two-dimensional solutions is more favorable with respect to $T_C$ from the experimental point of view. The most important experimental results, imposing limits on the theoretical estimates were obtained for 2D $^3$He submonolayers on the surface of thin $^4$He films by Pobell [11.14] and on the surface of grafoil by Saunders, Neki et al [11.13, 11.63]. The authors of [11.14] performed the measurements of the viscous penetration length with the aid of torsional oscillations. The scheme of their experiment is shown on Fig. 11.5.

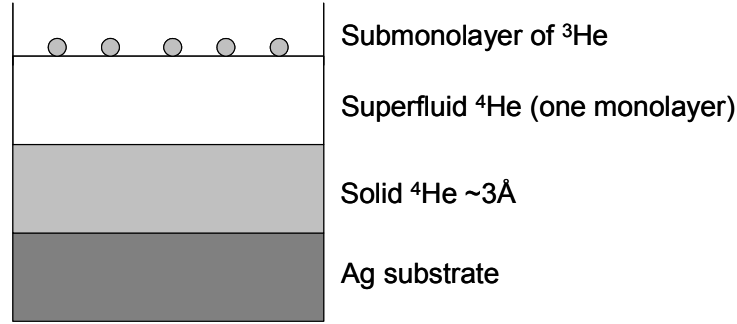

Fig. 11.5. The principal scheme of Pobell experiments [11.14, 11.6] on the search of fermionic superfluidity in the submonolayers of $^3$He on the surface of very thin film of superfluid $^4$He.

Pobell go down to the temperatures of 0.85 mK and did not observe the superfluid transition in the range of surface concentrations from 0.1 to 1 monolayer. Even more severe restrictions on $T_C$ impose the experiments of Saunders group [11.13, 11.63] who did not find the fermionic superfluid transition up to the temperatures 0.1 mK for $^3$He surface coverages from 0.05 till 1 monolayer on the grafoil substrate.

These experimental data suggest that the fermionic superfluidity in two-dimensional solution should be searched either at lower $^3$He densities ($n_3 < 0.03$ of a monolayer, when the total interaction corresponds to the attraction) or at a little bit higher densities $n_3 \sim 0.05 \div 1$ of a monolayer (for repulsive total interaction) when we apply a strong magnetic field $H \sim 15 \div 30$ T. In both cases we can probably expect the singlet superfluidity in the absence of a magnetic field (for concentrations $n_3 < 0.03$ of a monolayer) and the field-enhanced triplet superfluidity (for $n_3 \sim 0.05 \div 1$ of a monolayer) at temperatures of the order of 1 mK or less. These temperatures are ten times higher than in 3D case and in principle can be achieved experimentally.

11.5. Two-dimensional monolayers as a bridge between superfluidity and high-$T_C$ superconductivity.

In conclusion of this Chapter let us stress again that the two-dimensional $^3$He submonolayers on the surface of thin $^4$He films and on grafoil are ideal two-dimensional systems for experimental verification of many currently popular theories of the normal and superconducting state of quasi-two-dimensional high-$T_C$ superconductors, including marginal Fermi-liquid theory of Varma et al [11.31] and Luttinger liquid theory of Anderson [11.32]. The most valuable experimental information which can impose limitations on different theories of high-$T_C$ SC is connected with the measurements of a magnetic susceptibility $\kappa = \kappa_0 \dfrac{1 + \frac{1}{2} F_1^s}{1 + F_0^a}$ at low temperatures $T < T_F$ and small densities. The knowledge of Landau harmonics $F_0^a$, $F_1^s$, which enter in the expression for $\kappa$, will help us to answer the question whether Landau Fermi-liquid theory exists in a 2D case even at low densities, and, if it exists, what are the non-trivial corrections to Landau expansion for measurable quantities such as effective mass or zero-sound velocity, which are missing in a standard 3D situation. We will address this topic more detaily in Chapter 16, when we will study the singularity in Landau quasiparticle interaction function $f(\vec{p}, \vec{p}')$ for almost parallel momenta $\vec{p} \parallel \vec{p}'$ and small transition momentum $\vec{q} = \vec{p} - \vec{p}' \to 0$ [11.11, 11.12] in 2D and consider detaily the doubts of Anderson [11.32] on the validity of Landau Fermi-liquid description due to the existence of this singularity. We will show in Chapter 16 that in the framework of perturbation theory [11.11, 11.12] the singularity is much weaker $f(\vec{p}, \vec{p}') \sim \frac{1}{\sqrt{q}}$ then Anderson's prediction $f(\vec{p}, \vec{p}') \sim \frac{1}{q}$. Moreover it exists only for a narrow angle region $\varphi \sim q^{3/2}$ close to the angle $\varphi = 0$, thus not destroying Landau Fermi-liquid completely but leading only to nontrivial temperature corrections to the effective mass $(m^*/m - 1) \sim 1/2 F_1^s \sim f_0^2 T / \varepsilon_F$ and other observables. It will be very interesting to check experimentally the predictions of the perturbation theory of Prokof'ev, Stamp [11.11], Baranov, Kagan, Mar'enko [11.12] on the $^3$He submonolayers at low temperatures and surface densities. To do that it is important to expand the susceptibility measurements of Hallock and Godfrin [11.5] groups for $^3$He submonolayers on the surface of thin film of superfluid $^4$He and Saunders group for $^3$He submonolayers on grafoil on lower temperatures and surface densities where we are in degenerate (Pauli) situation. Note that earlier experiments [11.4, 11.44, 11.45, 11.17, 11.59] have been carried out primarily at intermediate and high temperatures (at which there is a transition to Curee law).

Reference list to Chapter 11.

Chapter 13. Spin-charge separation and confinement in ladder systems and in high-$T_C$ superconductors.

In the present Chapter on the level of 2D isotropic t-J model [13.1-13.3] and strongly anisotropic (quasi-1D) t-J model [13.4.-13.6] we will describe the superconductive phase diagram of layered cuprates and quasi-1D ladder materials [13.5, 13.33. 13.34]. We present the basic ideas of the physics of spin-charge separation [13.2, 13.8-13.10, 13.31, 13.40] and spin-charge confinement [13.21, 13.22]. Note that spin-charge separation is actual for 1D AFM spin-chains [13.8-13.10, 13.35-13.37] and extended regions of the phase diagrams of ladder systems with odd number of "legs" (odd number of coupled spin-chains such as three-leg ladders [13.11, 13.12], five-leg ladders and so on). Note also that in terminology used for ladder systems (see [13.5] for a review) each spin-chain, is called the "leg", while interchain AFM-coupling and hopping (in doped case) are described in terms of "rungs" (see Fig. 13.1).

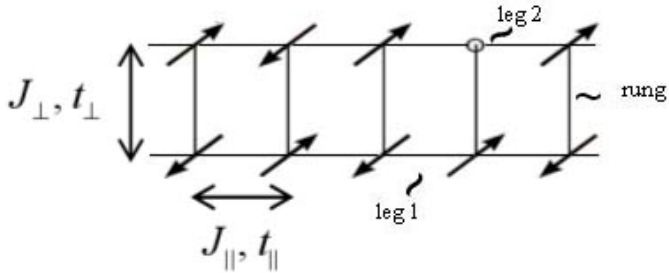

Fig. 13.1. A typical example of the two-leg ladder (a coupled system of two spin-chains). The interchain AFM-coupling and hopping (if at least one leg is doped) $J_\perp, t_\perp$ are described in terms of rungs. $J_\parallel, t_\parallel$ are AFM-coupling and hopping along the legs.

The system with spin-charge separation are usually described in terms of Luttinger Fermi-liquid (LL) [13.8.-13.10]. For LL the group velocities of spinons $v_S$ (spin excitations) and holons $v_C$ (charge excitations) are different [13.40], so the charge transfer (charge-density waves CDW) and spin transfer (spin-density waves SDW) are also described by different group velocities $v_S \neq v_C$. This phenomena can be better understood if we use electroneutrality considerations. They are important for charge transfer and so the crystalline lattice (the ions) also participate in the process. However we can neglect the ions for spin transfer. Thus we can understand the difference between the group velocities of CDW and SDW in 1D systems. Correspondingly in LL spin excitations (spin waves) are gapless Goldstone modes while charge excitations can be gapped or gapless depending upon the model (the finite mass can be generated for charge excitations in analogy with plasmons in metals). This leads to slow (power-law) decay of spin-spin correlations, while charge-charge correlations can be even rapidly exponentially decaying in 1D. If we, vice a versa, consider the ladders with even number of legs (characteristic example of two-leg ladder [13.13] is presented on Fig. 13.1), then we will fall in another universality class as people call it. Namely the two-leg ladder will be described by quite different Luther-Emery (LE) liquid [13.9, 13.38, 13.39]. In LE liquid we have opposite situation: the spin-spin correlations are exponentially decaying while density-density (and SC gaps) correlations are slowly decaying. In LL the dominant instability is towards SDW-formation, while in LE liquid it is towards CDW-formation and strong SC-fluctuations which favor SC if we include small interaction between ladders (see [13.13, 13.27]). In the limit of strong-coupling along the rungs $J_\perp \gg \{J_\parallel, t_\parallel, t_\perp\}$ we will show (see Fig 13.2) that for two holes it is energetically beneficial to occupy the same rung, thus forming a rung-boson (or biholon or local Cooper pair) with charge $2e$.

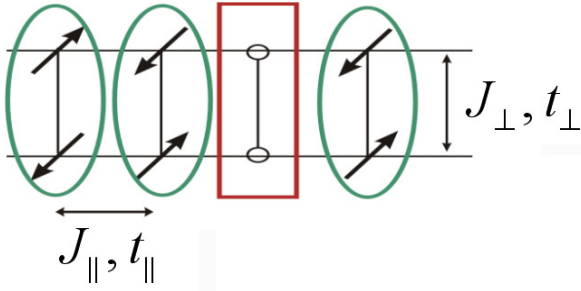

Fig 13.2. The bound state of two holes on one rung in the strong-coupling limit of the anisotropic t-J model for the two-leg ladder $J_\perp \gg \{J_\|, t_\|, t_\perp\}$. In this limit the spins on the rung form local Kondo-singlets.

In the same time in this limit the spins on the rung also form local Kondo-singlets with the $\Psi$-function $|\Psi_s\rangle = \frac{1}{\sqrt{2}}|\uparrow\downarrow - \downarrow\uparrow\rangle$ with an energy $E_s = -\frac{3}{4}J_\perp$. We can say that the rung-boson (or biholon or a local Cooper pair) moves in this limit in the surrounding of local (rung) Kondo-singlets [13.13]. Kondo-singlets play the pole of holes in effective 1D-model for the rungs. The effective 1D-model for this case is equivalent to 1D Bose-Hubbard model for rung-bosons [13.14]. In general we can say that LE-liquid is equivalent to 1D repulsive Bose-gas of composite (rung) bosons, while LL is equivalent (for the three-leg ladder for example) to 1D repulsive Fermi-gas or repulsive 1D Hubbard model of composite (rung) fermions [13.27, 13.28]. When we increase the number of legs (starting with three-leg ladders) the phase diagrams become more sophisticated. They contain both regions of LL and LE-liquid depending upon the relation between the parameters $\{J_\perp, t_\perp\}$ and $\{J_\|, t_\|\}$ describing AFM-coupling and hopping along the rungs and legs, respectively [13.11, 13.12]. The difference between even- and odd-numbers of legs becomes less pronounced when we increase the number of legs but still there is no smooth transition to the limit when $N$ (number of legs) $\to \infty$. However in the isotropic limit of the model $J = J_\perp = J_\|$, $t = t_\perp = t_\|$ and $J/t \sim \frac{1}{2} \div \frac{1}{3}$ (typical for 2D high-$T_C$ materials) there is a tendency towards the coexistence of LL and LE-liquids. Thus there is a motive of the Fermi-Bose mixture of effective (rung) fermions and bosons in the isotropic limit [13.15]. Correspondingly we can consider underdoped limit of 2D high-$T_C$ materials in the framework of a strongly interacting Fermi-Bose mixture of spinons and holons [13.16]. Note that the strong interaction between spinons and holons in the underdoped cuprates can be connected with linear (confinement) potential describing Bulaevskii-Nagaev-Khomskii [13.1] Brinkman-Rice [13.17] AFM-string [13.25]. It is easier however to proceed at first to more simple overdoped limit of 2D isotropic t-J model (see the beginning of Section 13.4).

In the second part of the Chapter we consider isotropic 2D t-J model [13.1-13.3]. At first we analyze superconductive phase diagram of this model in strongly overdoped case (for small and intermediate electron densities where Landau Fermi-liquid picture is valid). We find the regions of extended s-wave, p-wave and $d_{x^2-y^2}$-wave pairings as well as the tendency towards phase separation at large values of $J/t$ and low density [13.18-13.20]. We present then the simple estimates for d-wave critical temperature in the optimally doped case [13.19] and get reasonable $T_C \sim 100$ K typical for cuprates here [13.91, 13.26, 13.49]. Then we return back to difficult corner of the phase diagram of the t-J model for which $n_{el} = 1 - x \to 1$ ($x \ll 1$ - underdoped case) and $J/t \sim \frac{1}{2} \div \frac{1}{3}$. Here in agreement with Fermi-Bose mixture ideas [13.15] introduced in the first part of the Chapter, we propose a scenario of BCS-BEC crossover for pairing of two composite holes [13.21, 13.22] (two strings [13.25], or two spin-polarons [13.23, 13.24]) in the d-wave channel [13.16]. Note that each composite hole (each string or spin polaron) contains spinon and holon interacting via confinement potential. Here we are inspired by the ideas of Laughlin on spin-charge confinement [13.21, 13.22] and the analogies between composite holes in underdoped state of the 2D t-J model and quark-gluon physics (physics of quark bags) in quantum chromodynamics (QCD) [13.50-13.54]. (Note that alternative slave-boson [13.43 – 13.45, 13.48] spin-charge separation scenario was considered briefly (see also [13.57]) in connection with 2D underdoped t-J model in Chapter 5).

In the end of the Chapter, we briefly discuss the possible BCS-BEC crossover scenario for high-$T_C$ materials.

### 13.1. Spin-charge separation and Luttinger liquid in doped spin-chains.

We start the present Section with a brief enumeration of the powerful analytical methods developed in 1D physics (usually they are not so effective for higher dimensionalities).

They include Bethe-ansatz (Bethe [13.35]) for a 1D chain of spins $S = ½$, exact solutions in 1D Hubbard model for arbitrary density $n$ and coupling strength $U/t$ (Lieb, Wu [13.36]), nonlinear sigma model with the topological term for halfinteger and integer spins and methods of conformal field theory in 1 space + 1 time dimensions (Haldane [13.8], Belavin, Polyakov, Zamolodchikov [13.94], Pruisken [13.98], Frehm, Korelin [13.95], Hawakami, Yang [13.97]) as well as bosonization methods (Tonomaga [13.40], Luttinger [13.41]). While the exact solutions (see also Kawakami et al [13.96, 13.97]) are very useful to describe the ground state of 1D Hubbard model or 1D interacting Fermi-gas at different coupling strengths, the bosonization methods help to represent the energy of the low-lying excited states of the 1D interacting Fermi-systems as a sum of the energies of the independent bosonic oscillators.

We can say that to some extent bosonization method is ideologically similar to the hydrodynamical method presented in the first part of the manuscript. Effectively it is based on the introduction of the collective bosonic variables describing charge and spin-density fluctuations. However bosonization methods could give slightly more than hydrodynamics since in some models (such as Tomonaga-Luttinger model for example) they describe not only oscillations with small frequencies and $k$-vectors, but also help to restore density-density and spin-spin correlations at large $k$-vectors of the order of $2p_F$ related to giant 1D Kohn's anomaly (see also Chapter 9 for more details).

For the introduction to the bosonization method we can recommend an excellent review-article of Brazovskii and Kirova [13.93], pioneering articles [13.85-13.88] on abelian and nonabelian bosonization and textbooks [13.53, 13.54, 13.84] of Fradkin, Tsvelik et al., and Gimarchi.

We are much more modest in the present Section and will study mostly a 1D doped spin-chain with AFM-interaction between nearest neighbor spins $S = ½$ in the framework of the 1D t-J model.

#### 13.1.1. 1D t-J model for doped spin-chains.

Let us consider 1D t-J model for doped spin-chains with AFM interaction ($J > 0$) between spins $S = ½$. In the absence of doping the Hamiltonian is of the Heisenberg type and reads:

$$\hat{H} = J \sum_{<ij>} \vec{S}_i \vec{S}_j \qquad (13.1.1)$$

It is well known that in 1D spin-fluctuations destroy long-range AFM ordering. Spin excitations are gapless. Spin correlations decay in a power fashion in this model:

$$<\vec{S}(x)\vec{S}(0)> \sim \frac{\cos 2p_F x}{r^{1+\beta}}, \qquad (13.1.2)$$

where $\beta$ is model-dependent parameter. For small $J$ the parameter $\beta = ½$ and $1 + \beta = 3/2$. Note also that exact solution for 1D spin-chain is available for $S = ½$: $E = - J \ln2\, N$, where $N$ is the number of sites (see Bethe [13.35]). At low doping $n = 1 – x$ ($x \ll 1$) the system is described by the 1D t-J model [13.36, 13.37]:

$$\hat{H} = -t\sum_{<ij>} c^+_{i\sigma} c_{j\sigma} + JS \sum_{<ij>} \vec{S}_i \vec{S}_j \qquad (13.1.3)$$

It is shown in [13.29] that for $J < 2t$ the 1D t-J model [13.36, 13.30] belongs to the universality class of Luttinger liquid (LL). The same universality class as we mentioned already describes 1D repulsive Hubbard model and 1D Fermi-gas with repulsion. The basic instability in 1D t-J model for $J < 2t$ is with respect to spin-density wave (SDW).

13.1.2. Spin-charge separation in doped 1D spin-chains.

As we already mentioned, one of the most important features of LL is a phenomenon of phase separation. Let us illustrate this phenomenon for doped spin-chain, described by 1D t-J model, following Fulde (see Fig. 13.3 and [13.31]).

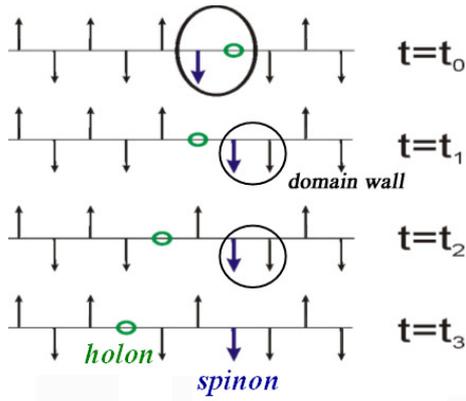

Fig. 13.3. Illustration of spin-charge separation in 1D according to Fulde (in "Strong correlations in molecules and Solids" [13.31]). If in the initial moment $t = t_0$ spinon and holon are nearby, than in the final moment $t = t_3$ there is a final distance $R = (v_S - v_C)\Delta t$ between them.

Qualitatively we can describe Fig. 13.3 in the following manner: in the initial moment $t_0$ spinon and holon are nearby. The holon starts to move on the left hand side. As a result the domain wall of two frustrated spins (in Ising limit) is created at $t = t_1$ [13.50, 13.54]. Finally in the moment $t = t_3$ the holon is separated from the domain wall by a regular structure of non-frustrated spins (in Ising limit). The distance between spinon and holon is $R = (v_S - v_C)\Delta t$. Note that, as we discussed in the Introduction to this Chapter, spinon and holon possess the different group velocities $v_S \neq v_C$ due to the fact that in charge transfer (charge-density wave or CDW) participate both electrons and ions because of electroneutrality. In the same time in spin transfer (SDW) participate only electrons. Of course, presented picture of spin-charge separation is oversimplified. When we are speaking about domain wall, for example, we should understand what approximation (slave-boson [13.42, 13.48] according to which spinon is fermion $f_{i\sigma}$ with charge 0 and spin ½ while holon is boson $b_i$ with charge $e$ and spin 0, or slave-fermion [13.46, 13.47] where vice a versa holon is spinless fermion $h_i$, while spinon is $S = 1$ boson $S_i^+$ ($S_i^-$) (a bit similar to magnon) is better. It seems that slave-fermion approximation is more reliable here [13.55] since a domain-wall corresponds with some degree of precaution to a localized magnon.

13.1.3. The dressed Green-function in 1D Luttinger liquid.

The dressed Green-function in standard Landau Fermi-liquid has a simple one-pole structure close to the Fermi-surface (for $p \to p_F$). Correspondingly we get in 3D or 2D Fermi-gas: $G(\omega, \vec{p}) = \dfrac{Z}{\omega - \xi_p + i\delta}$, where $\xi_p = \dfrac{p^2 - p_F^2}{2m}$ is uncorrelated quasiparticle spectrum and $Z$ is quasiparticle residue or Z-factor. However in 1D systems, which are described by Luttinger liquid and are subject of spin-charge separation, the situation is drastically changed. The dressed Green-function does not have a simple one-pole structure close to the Fermi-surface. Instead of that it often has a brunch-cut in momentum space for p ∈ [-$p_F$, $p_F$]. In the most simple Tomonaga model for 1D spinless fermions, for example, according to Dzyaloshinskii, Larkin [13.32]:

$$G(\omega, p) \sim \frac{Z(\omega)}{\sqrt{[\omega - v_F(|p| - p_F)][\omega - u(|p| - p_F)]}}, \quad (13.1.4)$$

where Z-factor is vanishing on the Fermi-surface:

$$Z(\omega) \sim \omega^\alpha \to 0 \text{ for } \omega \to 0 \text{ (or for } |p| = p_F), \quad (13.1.5)$$

and the power $\alpha > 0$ depends upon the model.

Let us emphasize that the Green-function has a typical square-root in denominator of (13.1.4) and moreover the velocity $u \neq v_F$. If we include spin degrees of freedom, the expression for the dressed one-particle Green-function $G(\omega, p)$ become rather cumbersome in momentum space (see Ren, Anderson [13.90] and Medden, Schonhammer [13.91]) containing a hypergeometric function.

However is real space it has a typical for spin-charge separation square-root again (see [13.53, 13.92]). If we make Wick transformation $t \to i\tau$ and introduce conformal variables $z = \tau + \frac{ix}{v}$ and $\bar{z} = \tau - \frac{ix}{v}$, then the Green-function $G(x, \tau)$ in Tonomaga-Luttinger model reads in weak-coupling case:

$$G(x, \tau) = \left(\frac{a^2}{v_s^2 \tau^2 + x^2}\right)^\alpha \left[\frac{\exp(ik_F x)}{\sqrt{(v_s \tau - ix)(v_c \tau - ix)}} + \frac{\exp(-ik_F x)}{\sqrt{(v_s \tau + ix)(v_c \tau + ix)}}\right],$$

where $0 < \alpha < 1/8$ is model-dependent parameter and $\alpha = 1/8$ in strong-coupling limit (of the Hubbard model). Of course the velocities $v_C$ and $v_s$ are different. In weak-coupling limit of Tonomaga-Luttinger or 1D Hubbard model $v_C = v - g/2\pi$ and $v_s = = v + g/2\pi$, where $g = U/4t < 1$ is a coupling constant. In strong-coupling limit of 1D Hubbard model $v_C/v_s \sim 1/g << 1$. Thus for short-range repulsive interaction in 1D both spinons and holons are gapless. The charge excitations could be gapped in LL in case of long-range repulsive interaction. Finally let us repeat that if we have for example a two-leg ladder (a coupled system of the two spin chains), then the system falls in another universality class of Luther-Emery (LE) liquid. As we already mentioned LE liquid is equivalent to 1D Fermi-gas with <u>strong</u> attraction or to 1D Bose-gas with repulsion. In LE liquid we always have a gapless (Bogolubov) spectrum for charge excitations. In fact in this case we have a sound wave in 1D Bogolubov Bose-gas. The spectrum for spin excitations is gapped. We can say that while in LL the basic instability is towards spin-density wave (SDW) formation, in LE liquid the basic instability is towards charge-density wave (CDW) formation and strong superconductive (SC) fluctuations. LL describes spin-chains with half-integer spins and extended regions of phase-diagrams of odd-leg ladders (see the discussion of three-leg ladders in the next Sections). LE liquid describes spin-chains with integer spins (see Haldane [13.8]) and extended regions of phase-diagrams of even-leg ladders. It is seductive to describe 2D high-$T_C$ materials as a Fermi-Bose mixture of strongly interacting LL and LE liquids (see the next Section).

13.1.4. The distribution function for interacting particles in Luttinger liquid.

The interacting particles (not quasiparticles) distribution function $N_{int}(p) = \int \frac{d\omega}{2\pi} G(\omega, p)$ with $G(\omega, p)$ being a dressed one-particle Green-function from (13.1.4) does not have a finite jump on the Fermi-surface [13.8-13.10, 13.33, 13.81-13.82]: $Z \sim \to 0$ for $\omega \to 0$ (see (13.1.5) for Luttinger liquid (LL)). Instead of the jump $N_{int}(p)$ possesses the power-law singularity close to $p_F$ which reads (see Lieb, Mattias [13.82], Fulde [13.3] and Fig.13.4):

$$N_{int}(p) = \frac{1}{2} - const |p - p_F|^\alpha \, sign(p - p_F), \qquad (13.1.6)$$

The exponent $\alpha$ depends upon the model. In strong-coupling limit of 1D Hubbard model $U \gg t$ (which close to half-filling is practically equivalent to t-J model with $J = \frac{4t^2}{U} > 0$ yielding the same energy $E = -J \ln 2 \, N$ exactly at half-filling) this constant $\alpha = 1/8$.

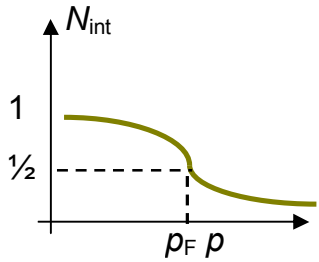

Fig.13.4. Interacting particles distribution function in LL. There is no jump at $p = p_F$. Instead of it there is power-law singularity (see [13.31]).

For comparison on Fig.13.5 we present the interacting particles distribution function for Landau Fermi-liquid (LFL) with a finite jump ($Z \neq 0$) for $p = p_F$ [13.42].

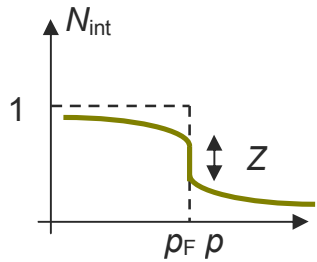

Fig.13.5. Interacting particles distribution function in Landau Fermi-liquid with a finite jump ($Z \neq 0$) at $p = p_F$ (see [13.31]).

Note also that the imaginary part of the dressed (by interactions) one-particle Green-function in Landau Fermi-liquid has a sharp quasiparticle $\delta$-functional peak for $\omega = \xi_p = \varepsilon(p) - \mu$ and a broad incoherent background (see Fig.13.6).

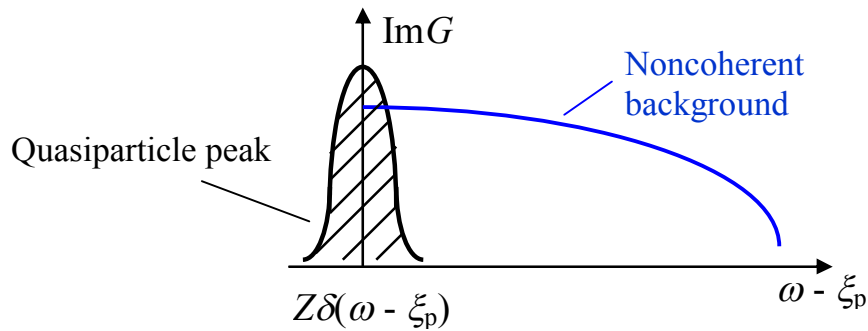

Fig.13.6. Imaginary part of the dressed one-particle Green-function in Landau Fermi-liquid theory. There is a sharp $\delta$-functional quasiparticle peak and a broad incoherent background on it.

In contrast to situation on Fig.13.6 in Luttinger liquid quasiparticle $\delta$-functional peak is absent ($Z = 0$) and we have only noncoherent background in imaginary part of the dressed one-particle Green-function. Note that in spite of these substantial differences both LFL and LL conserve the volume of the Fermi-sphere due to the Luttinger theorem [13.41, 13.42], which means that the number of interacting particles ($N_F$) equals to the number of free fermions in them ($n = p_F^3/6\pi^2$ in 3D case, $n = p_F^2/4\pi$ in 2D and $n = p_F/\pi$ in 1D for spinless particles).

It is interesting to stress that (as we will see later on) in the case of spin-charge confinement (typical for 2D high-$T_C$ compounds) the dressed one-particle Green-function (see Lee et al. [13.43-13.45]) reads:

$$G(\omega, \vec{p}) \sim \frac{Z(\omega)}{\omega - \xi(p) + io} + G_{incoh}(\omega, \vec{p}), \qquad (13.1.7)$$

where $\xi(p) = E(p) - \mu$ is a spectrum of a 2D AFM-string, $E(p) = E_0 + Jp^2$ is a string energy (see Subsection 13.4.8) [13.25]. Thus, the dressed one-particle Green-function for a composite hole (for a string) has a simple (one-pole) structure close to the Fermi-energy (for small $\xi(p)$) in the case of spin-charge confinement and corresponds to Landau Fermi-liquid.

13.2. Two-leg ladder systems. Spin-charge confinement. Luther-Emery liquid.

In the introduction to this Chapter we already illustrated the structure of the two-leg ladders on Figs. 13.1 and 13.2. Let us present several experimentally available examples of two-leg ladder materials $(VO)_2P_2O_7$, $LaCuO_{2.5}$, $SrCu_2O_3$, and $NaV_2O_5$ [13.5]. There is also one example of the ladder material $Sr_xCa_{14-x}Cu_{24}O_{41}$ [13.33, 13.34] where superconductivity was observed with critical temperatures $T_C \div (9 - 12)$ K for pressures $P \sim (3 \div 4)$ GPa. Note that in $NaV_2O_5$ a strongly anisotropic case $J_\perp \sim 4J_\parallel$ is realized. On Fig 13.7 we present typical in-plane situation for two-leg ladder materials. It is instructive also to present a crystalline structure of a typical two-leg ladder material $SrCu_2O_3$ (see Fig.13.8).

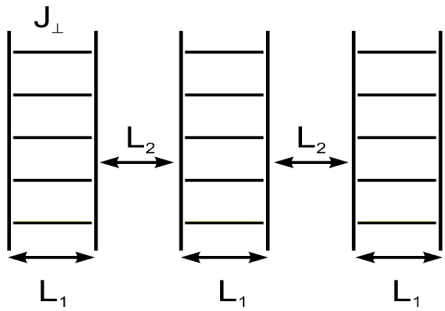

Fig 13.7. In-plane situation for two-leg ladder materials.

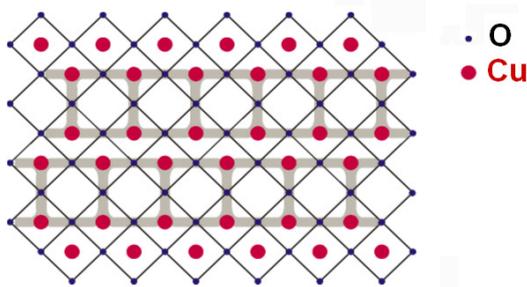

Fig.13.8. Crystalline structure of a typical two-leg ladder material $SrCu_2O_3$. In the center of each elementary $CuO_4$ plaquette there is Cu atom (red circle), while in the corners of the plaquette there are four O-atoms (see [13.5]).

Note that in contrast to stripes in high-$T_C$ materials [13.18, 13.58] the ladders are stable (strong) defects of the crystalline lattice which are not fluctuating and exist even in the absence of doping. According to Goodenough rules [13.59] for the chemical bond we have strong coupling (AFM superexchange) inside the ladder corresponding to the Cu-O-Cu bond angle $\pi$ (see Fig. 13.9) and weak coupling of FM-type on the bond (with Cu-O-Cu angle $\pi/2$) between the ladders.

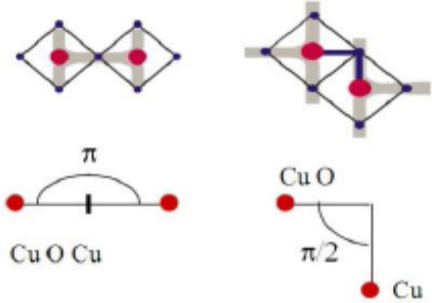

Fig. 13.9. Strong coupling AFM superexchange with an angle $\pi$ for Cu-O-Cu bond and weak coupling FM Cu-O-Cu bond (with an angle $\pi/2$) between the ladders.

It is important that for two-leg ladder spin susceptibility acquires a gap (see [13.5] and Fig. 13.10):

$$\chi(T) \sim \frac{1}{\sqrt{T}} e^{-\Delta/T} \qquad (13.2.1)$$

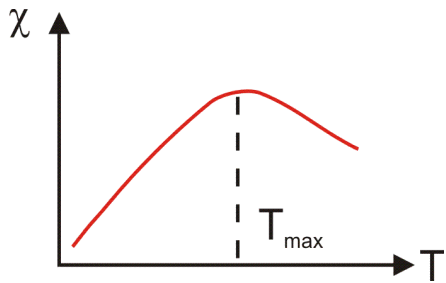

Fig. 13.10. Spin susceptibility for two-leg ladder materials.

From Fig. 13.10. we see that $\Delta \sim T_{max}$ and moreover for $T \ll T_{max}$: $\chi(T) \to 0$.

13.2.1. Anisotropic t-J model.

In the absence of doping the two-leg ladders are described by anisotropic Heisenberg Hamiltonian:

$$\hat{H} = J_\| \sum_{ia} \vec{S}_{ia} \vec{S}_{i+1a} + J_\perp \sum_{ia} \vec{S}_{ia} \vec{S}_{ia+1} \qquad (13.2.2)$$

where $J_\|$ is AFM-exchange along the legs and $J_\perp$ - along the rungs (see Fig. 13.2), the index $i$ is a rung index and index $a = 1, 2$ corresponds to the leg 1 and leg 2. For $J_\perp \gg J_\|$ the spin-singlets are formed on the each rung (see Fig. 13.11). The ground state for $J_\| = 0$ corresponds to dimerized spin-liquid with $S_{tot} = 0$. The total energy in this case $E_s = -\frac{3}{4} J_\perp N_{rung}$ and the $\Psi$-function

$$\Psi_s = \prod_i |\Psi_i\rangle, \text{ where } |\Psi_i\rangle = \frac{1}{\sqrt{2}} |\uparrow\downarrow - \downarrow\uparrow\rangle.$$

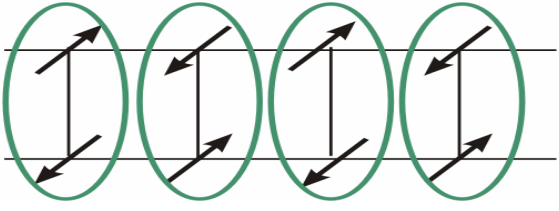

Fig. 13.11. Local Kondo-singlets in the strong coupling $J_\perp \gg J_\parallel$ of the anisotropic Heisenberg model for two-leg ladders in the absence of doping.

Spin-gap $\Delta$ in the magnetic susceptibility is in fact the difference between singlet and triplet energies: $\Delta = E_t - E_S$. For $J_\perp \ll J_\parallel$: $\Delta \sim \exp\{-1/J_\perp\}$ (see [13.13]). For $J_\perp = J_\parallel$ (isotopic point): $\Delta = 0.5 J_\perp$. Finally for $J_\perp \gg J_\parallel$: $\Delta \to J_\perp$. The last result is evident since for $J_\perp \gg J_\parallel$: $E_t = 1/4\, J_\perp$, while $E_S = -3/4 J_\perp$. Note that the spin-gap opens already at $J_\perp \to 0$ (see Fig. 13.12 and [13.13]). Thus $J_\perp = 0$ is a singular point.

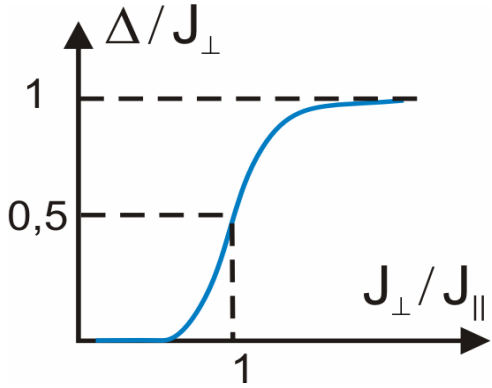

Fig. 13.12. Spin-gap in anisotropic Heisenberg model (see [13.5, 13.6]).

Spin correlator acquires an exponential form:

$$\langle \vec{S}(r)\vec{S}(0) \rangle \sim \exp\left(-\frac{r}{\xi}\right), \quad (13.2.3)$$

where the correlation length $\xi = \dfrac{v_s}{\Delta}$. In the limit of low-doping two-leg ladder is described by anisotropic t-J model (see Fig.13.2):

$$\hat{H} = -t_\parallel \sum_{ia\sigma} c^+_{ia\sigma} c_{i+1a\sigma} - t_\perp \sum_{ia\sigma} c^+_{ia\sigma} c_{ia+1\sigma} + J_\parallel \sum_{ia} \vec{S}_{ia}\vec{S}_{i+1a} + J_\perp \sum_{ia} \vec{S}_{ia}\vec{S}_{ia+1} \quad (13.2.4)$$

where $t_\parallel$ and $t_\perp$ are hopping along the legs and along the rungs, respectively, $c_{ia\sigma}^+$ corresponds to the creation of electron on rung $i$ and leg $a$ with spin-projection $\sigma$. In the strong coupling $J_\perp \gg \{J_\parallel, t_\parallel, t_\perp\}$ it is more energetically beneficial for two holes to create a bound state on the rung (biholon or local Cooper pair (see Fig.13.2)). Biholon moves in the surrounding of spin-singlets. Its effective mass reads:

$$m_{eff} \sim \left(\frac{1}{t_\parallel d^2}\right)\frac{J_\perp}{t_\parallel} \gg \frac{1}{t_\parallel d^2}. \quad (13.2.5)$$

We have 1D Bose-gas with repulsion for bosons (biholons) on the rungs. It belongs to the universality – class of Luther-Emery (LE) liquid. In this type of model there are strong SC fluctuations:

$$\langle \Delta(r)\Delta(0) \rangle \sim \frac{1}{r^{1+\gamma}} \quad (13.2.6)$$

Note that according to Efetov, Larkin [13.14] already a small interaction between the ladders (see Figs.13.7 and 13.8) stabilizes a finite $T_C$ in the system.

### 13.2.2. Resistivity in two-leg ladders materials.

For the material $La_{1-x}Sr_xCuO_{2.5}$ with two-led ladders resistivity behaves as follows (see Fig.13.13).

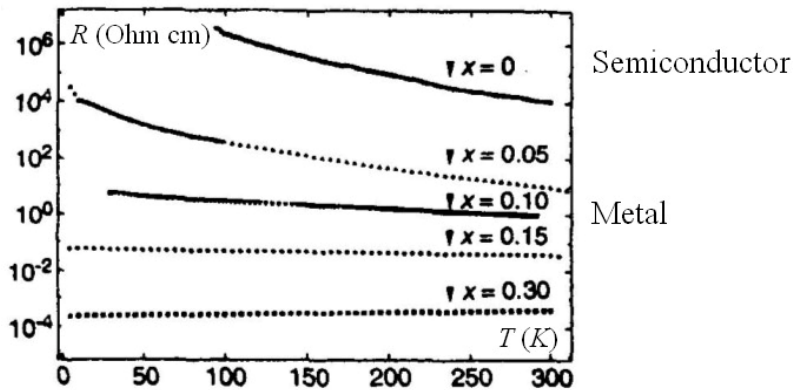

Fig.13.13. Resistivity characteristic $R(T)$ in the two-leg ladders material $La_{1-x}Sr_xCuO_{2.5}$ for different doping levels (see [13.5]).

From Fig.13.13 we can see that for $x \geq 0.15$ resistivity $R(T)$ behaves in a metallic fashion in analogy with high-$T_C$ materials. At small doping $x \leq 0.10$ the resistivity behaves in a semiconductor fashion.

### 13.2.3. Superconductivity in ladder materials.

Superconductivity was experimentally observed in $Sr_xCa_{14-x}Cu_{24}O_{41}$ (see [13.33, 13.34]). In this compound in analogy with high-$T_C$ material YBaCuO there are chains and planes. We have two-leg ladders in planes (see Fig.13.14).

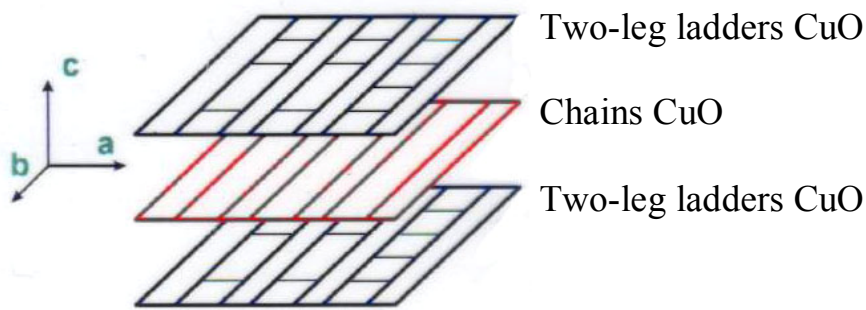

Fig.13.14. The crystalline "sandwich" structure of superconductive material $Sr_xCa_{14-x}Cu_{24}O_{41}$. There are two-leg ladders in two adjacent CuO planes and CuO chains in between (see [13.33, 13.34]).

For $x = 0.4$ and pressures $P < 3$ GPa the holes mostly occupy the chains. Resistivity has a semiconductive character (see Fig. 13.15). For $x = 0.4$ and pressures $3$ GPa $< P < 4.5$ GPa the holes mostly occupy the ladders. Resistivity behaves in a metallic fashion $\rho \sim \rho_0 + AT^\alpha$ with $1 < \alpha < 2$. Here at $T < T_C$ ($T_C = 12$ K for $P = 3$ GPa and $T_C = 9$ K for $P = 4.5$ GPa) SC arises in the system [13.5].

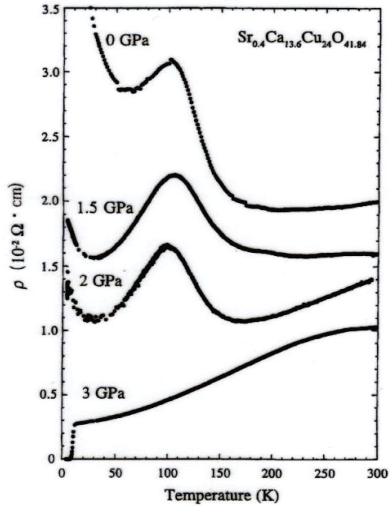

Fig.13.15. Resistivity characteristics $R(T)$ in the material $Sr_xCa_{14-x}Cu_{24}O_{41}$. For $P = 3$ GPa we observe the SC-transition $T_C = 12$ K see [13.33, 13.34]).

Intermediate conclusions for SC in $Sr_xCa_{14-x}Cu_{24}O_{41}$.

Let us emphasize once more that:
1) for $x = 0.4$ and $P > 3$ GPa the lattice is compressed and hence the holes mostly occupy the planes which contain two-leg ladders. This fact leads to metallic behavior of resistivity. As a result SC arises in the system.
2) For further increase of hole-concentration $x$ we will have an additional transfer of hole states from chains to planes.
3) For the first time SC was observed in $Sr_xCa_{14-x}Cu_{24}O_{41}$ for $x = 0.2$. In these materials SC arises for $P > 2.6$ GPa. The critical temperature is $T_C = 5$ K. In this case the hole concentration corresponds to hole density of 0.2 holes per Cu-atom of the ladder.

13.3. Three-leg ladders. Anisotropic t-J model for strong-coupling along the rungs.

The typical examples of three-leg ladder materials are $Sr_2Cu_3O_5$ and $CsCuCl_3$ (see Fig. 13.16 and [13.33, 13.34]).

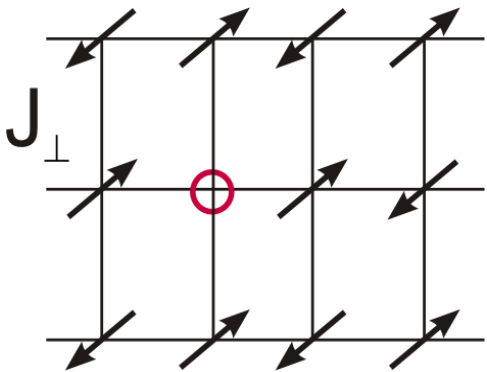

Fig. 13.16. Doped three-leg ladder.

In the limit of not very strong exchange $J_\perp$ along the rungs the spin-gap in susceptibility $\chi(T)$ is absent (see Fig.13.17).

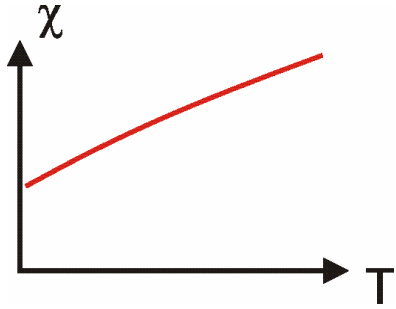

Fig.13.17. Spin-susceptibility $\chi(T)$ for three-leg ladders. $\chi(T) \to const$ for $T \to 0$ and the spin-gap is absent.

For low temperatures $T \to 0$ we can see from Fig.13.17 that $\chi(T) \to const$ and the spin-gap is absent. In the limit $\{J_\perp, t_\perp\} \gg \{J_\|, t_\|\}$ – we have strong coupling along the rungs [13.12]. In this limit the phase-diagram of three-leg ladders at low temperatures $T \to 0$ has extended regions of LL and LE-liquid for large $J_\perp/t_\perp$-ratio {see [13.12] and Fig.13.18).

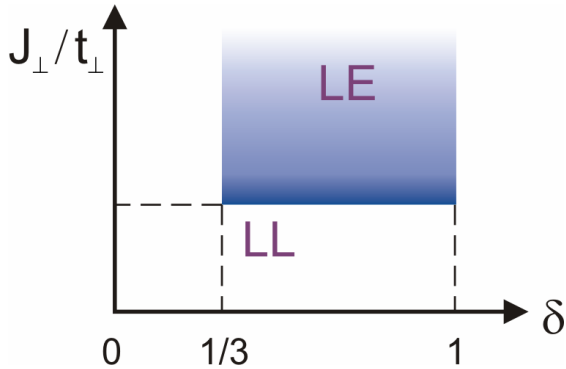

Fig.13.18 Phase-diagram of three-leg ladders at strong coupling along the rungs $\{J_\perp, t_\perp\} \gg \{J_\|, t_\|\}$. There are extended regions of LL and LE-liquid on the phase-diagram.

13.3.1. Exact diagonalization of one rung problem.

To construct more precisely the phase-diagram of three-leg ladder at strong coupling along the rungs $\{J_\perp, t_\perp\} \gg \{J_\|, t_\|\}$, we first should diagonalize (solve exactly) the one rung problem. Here in the limit $J_\| = t_\| = 0$ the $\Psi$-function of 3 spins (and zero holes) on the rung (see Fig.13.19) reads:

$$\Psi_0 = \frac{1}{\sqrt{6}}\left[|\uparrow\uparrow\downarrow\rangle - 2|\uparrow\downarrow\uparrow\rangle + |\downarrow\uparrow\uparrow\rangle\right] \qquad (13.3.1)$$

The $\Psi$-function $\Psi_0$ describes a spinon $f_{i\sigma}^+$ [13.2] with an energy $E_0 = -\frac{3}{2}J_\perp$, rung spin $S_{tot} = \frac{1}{2}$ and projection of rung spin $S_{tot}^z = \pm \frac{1}{2}$.

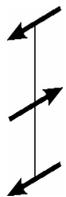

Fig.13.19. Spinon $f_{i\sigma}$ for three spins on the rung described by $\Psi$-function (13.3.1). It corresponds to rung spin $S_{tot} = \frac{1}{2}$ and $S_{tot}^z = \pm \frac{1}{2}$.

The Ψ-function of 2 spins and 1 hole on the rung (see Fig.13.20) corresponds to a holon $b_i^+$ and reads (see [13.12]):

$$\Psi_1 = \frac{1}{\sqrt{4+2\alpha^2}}\left[|\uparrow\downarrow 0\rangle - |\downarrow\uparrow 0\rangle + \alpha|\uparrow 0 \downarrow\rangle - \alpha|\downarrow 0 \uparrow\rangle + |0 \uparrow\downarrow\rangle - |0 \downarrow\uparrow\rangle\right] \quad (13.3.2)$$

The energy of this configuration is given by:

$$E_1 = -\frac{4t_\perp^2}{\sqrt{J_\perp^2 + 8t_\perp^2} - J_\perp} = -\frac{2t_\perp}{\alpha} \quad (13.3.3)$$

The total spin for this configuration $S_{tot} = 0$.

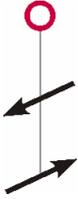

Fig.13.20. Holon $b_i^+$ for the rung with 2 spins and 1 hole. The total spin for this configuration $S_{tot} = 0$.

For 1 spin and 2 holes on the rung the Ψ-function reads:

$$\Psi_2 = \frac{1}{2}\left[|\uparrow 00\rangle + \sqrt{2}|0 \uparrow 0\rangle + |00 \uparrow\rangle\right] = \frac{1}{2}\left[c_{i1\uparrow}^+ + \sqrt{2}c_{i2\uparrow}^+ + c_{i3\uparrow}^+\right]|000\rangle \quad (13.3.4)$$

The energy of this configuration $E_2 = -\sqrt{2}t_\perp$. It corresponds to a spinon $h_{i\sigma}^+$ (Fig. 13.21) with $S_{tot} = \frac{1}{2}$ and $S^z_{tot} = \pm \frac{1}{2}$.

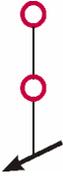

Fig. 13.21. Spinon $h_{i\sigma}^+$ for 1 spin and 2 holes on the rung. The total spin for this configuration $S_{tot} = \frac{1}{2}$ and $S^z_{tot} = \pm \frac{1}{2}$.

Finally for three holes on the rung the Ψ-function is trivial $\Psi_3 = |000\rangle$ and total energy $E_3 = 0$. It corresponds to a holon $a_i^+$ with $S_{tot} = 0$ (Fig.13.22).

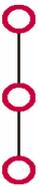

Fig.13.22. Holon $a_i^+$ for three holes on the rung. The total spin of this trivial configuration $S_{tot} = 0$.

### 13.3.2. Qualitative phase-diagram.

Now if we switch on small (but non-zero) $J_\parallel$ and $t_\parallel$ we get the following phase-diagram of three-leg ladder systems (see Fig.13.23). On Fig.13.23 LL I corresponds to the admixture of the rungs with 2 spins and 3 spins, LL II – to the admixture of the rungs with 2 spins and 1 spin, LL

III – to the admixture of the empty rungs and the rungs with 1 spin. In the same time LE-liquid corresponds to an admixture of the empty rungs and the rungs with 2 spins [13.12]. It is realized for doping $x > ⅓$ (see [13.12]):

$$\frac{J_\perp}{t_\perp} > \left(\frac{J_\perp}{t_\perp}\right)_{crit} = \frac{3}{\sqrt{2}}. \qquad (13.3.5)$$

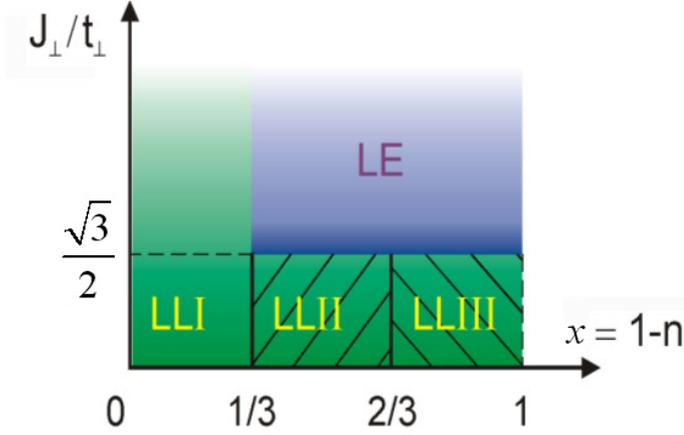

Fig.13.23. Phase-diagram of three-leg ladder systems at strong coupling along the rungs. There are three different regions of LL: LL I, LL II, and LL III (depending upon doping $x = 1 - n_{el}$) and a region of LE-liquid for larger values of $J_\perp/t_\perp$ and $x > ⅓$.

Let us consider the phase-diagram on Fig.13.23 more detaily. In the case of LL I when we include $t_\parallel \neq 0$ the hopping takes place due to an exchange between a rung with 3 spins and a rung with 2 spins (with a hole) (see Fig.13.24).

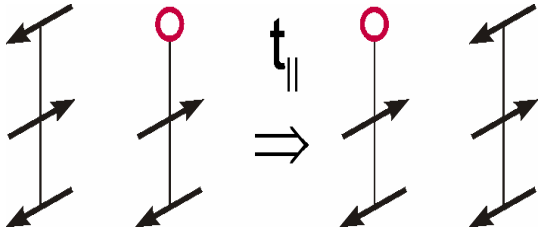

Fig.13.24. The hopping between a rung with 3 spins and a rung with 2 spins (with a hole) in case when $t_\parallel \neq 0$.

In this situation a composite fermion

$$g_{i\sigma}^+ = f_{i\sigma} b_i^+ \qquad (13.3.6)$$

arises in the problem. In the limit $\{J_\perp, t_\perp\} \gg \{J_\parallel, t_\parallel\}$ a composite fermion satisfies the standard fermionic anticommutational relations:

$$\{g_{i\sigma}^+, g_{i\sigma}\} = g_{i\sigma}^+ g_{i\sigma} + g_{i\sigma} g_{i\sigma}^+ = 1 \qquad (13.3.7)$$

Hence LL I corresponds to repulsive 1D Hubbard model for composite fermions $g_{i\sigma}$ described by the Hamiltonian:

$$\hat{H} = -t_{eff} \sum_{<ij>\sigma} g_{i\sigma}^+ g_{j\sigma} + U_\infty \sum_i n_{i,\sigma} n_{i,-\sigma}, \qquad (13.3.8)$$

where $U_\infty$ is infinity strong Hubbard repulsion on site $i$ between composite fermions. In the similar way we can understand LL II and LL III. The situation changes, however, for LE-liquid.

Here the hopping (for $t_\parallel \neq 0$) takes place due to an exchange between an empty rung and the rung with 2 spins (see Fig. 13.25).

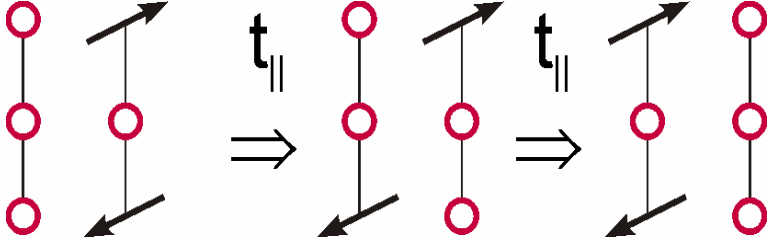

Fig. 13.25. The hopping (for $t_\parallel \neq 0$) between an empty rung and the rung with 2 spins.

It is easy to see that a composite boson
$$d_i^+ = a_i b_i^+ \qquad (13.3.9)$$
arises in the problem. Thus LE liquid corresponds to 1D Bose-gas with repulsion between composite bosons described by the Hamiltonian:
$$\hat{H} = -t_{eff} \sum_{<ij>} d_i^+ d_j + \frac{1}{2} U_\infty \sum_i n_i^2 . \qquad (13.3.10)$$

It is interesting to note that in contrast with (13.3.8) (where $t_{eff} \sim t$) in (13.3.10) $t_{eff} \sim t_\parallel^2 / J_\perp$ - appears only in the second order of perturbation theory.

### 13.3.3. $N$-leg ladders.

When we increase the number of legs the difference between the ladders with even and odd numbers of legs becomes less pronounced. For $N \to \infty$ the spin-gap in the ladders with even number of legs decreases exponentially. In strong coupling limit:
$$\Delta_{2N} \sim \frac{\Delta_2}{2^N} \sim \frac{J_\perp}{2^N} \to 0. \qquad (13.3.11)$$
Hence for $N \to \infty$ the spin excitations in the ladders with even number of legs are practically gapless (as in LL). Note that for $N \to \infty$ we proceed to two-dimensional anisotropic t-J model and for $\{J_\perp, t_\perp\} \gg \{J_\parallel, t_\parallel\}$ - to 1D $t_\perp$-$J_\perp$ model [13.12]. The universality class of this model corresponds to LL with spin-charge separation. This limit, however, is not realistic for 2D high-$T_C$ compounds.

### 13.3.4. The gap in the energy spectrum for three-leg ladders in anisotropic limit.

Returning back to three-leg ladders and their phase-diagram, we see that for doping $x > \frac{1}{3}$ and $J_\perp > \frac{3 t_\perp}{\sqrt{2}}$ there is an energy gap for $t_\perp > t_\parallel$. By the order of magnitude it reads:
$$\Delta = \frac{E_{LE} - E_{LL}}{N_{rung}} \sim t_\perp - t_\parallel . \qquad (13.3.12)$$
This gap separates LE and LL in energetic space.

### 13.3.5. Coexistence of bosonic Luther-Emery liquid and fermionic Luttinger liquid in isotopic limit.

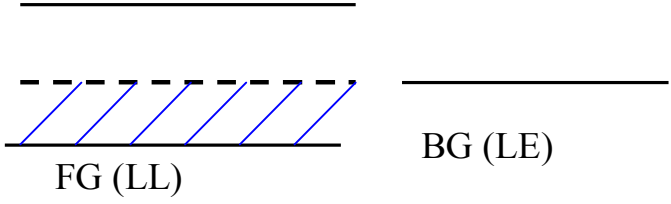

Fig. 13.26. The coexistence of Fermi (LL) and Bose-gas (LE) in the isotopic limit in the energy space.

The real high-$T_C$ materials correspond, however, to the difficult isotopic limit

$$J_\perp = J_\parallel = J; \quad t_\perp = t_\parallel = t; \quad J \sim 0.3t \qquad (13.3.13)$$

In this case the numerical calculations [13.11] show the tendency to a coexistence of Bose-gas (LE) and Fermi-gas (LL) in the energy space (see Fig. 13.26). In this case the energy gap

$$\Delta = \frac{E_{LE} - E_{LL}}{N_{rung}} \to 0 \qquad (13.3.14)$$

vanishes due to isotropic condition ($t_\perp = t_\parallel$) for hopping integrals. Hence in the isotopic limit we have Fermi-Bose mixture of LL and LE-liquids (see Geshkenbein, Ioffe, Larkin [13.15]).

### 13.3.6. Strongly interacting mixture of spinons and holons in high-$T_C$ superconductors.

We already mentioned briefly a Fermi-Bose mixture of spinons and holons in Chapter 5. Our project for underdoped high-$T_C$ superconductors reads: starting with a strongly interacting Fermi-Bose mixture of spinons $f_{i\sigma}^+$ and holons $b_i$ to derive an effective one-band model for the weakly interacting composite holes (or spin-polarons)

$$h_{i\sigma} = f_{i\sigma} b_i \qquad (13.3.15)$$

We will qualitatively consider this scenario in the last part of this Chapter on the basis Bulaevskii-Nagaev-Khomskii [13.1], Brinkman-Rice [13.17] ideas on AFM-string [13.2]. But at first we will understand the more simple overdoped limit of the isotopic t-J model in 2D case [13.18, 13.19], where Landau Fermi-liquid picture is valid and where we will have different SC instabilities, including d-wave pairing actual for real cuprates.

### 13.4. Superconductivity in isotropic 2D t-J model.

In this Section we will consider different superconductive pairings (s-wave, d-wave, p-wave) which arise in isotropic 2D t-J model in overdoped case, as well as a possible scenario of BCS-BEC crossover (or of a bosonic motive) which arises in an underdoped case of the t-J model.

### 13.4.1. Superconductive pairing in overdoped 2D t-J model.

In connection with high-$T_C$ superconductivity in the overdoped case we consider isotropic t-J model with released constraint [13.19]. The Hamiltonian of the model reads:

$$\hat{H} = -t \sum_{\langle ij\rangle\sigma} c_{i\sigma}^+ c_{j\sigma} + U \sum_i n_{i\uparrow} n_{i\downarrow} + J \sum_{\langle ij\rangle}\left(\vec{S}_i\vec{S}_j - \frac{1}{4}n_i n_j\right), \qquad (13.4.1)$$

where $n_{i\sigma} = c^+_{i\sigma} c_{i\sigma}$ is onsite density for spin projection σ, $\vec{S}_i = \frac{1}{2} c^+_{i\mu} \vec{\sigma}_{\mu\nu} c_{i\nu}$ is an operator of electron spin on site $i$, $\vec{\sigma} = \{\sigma_1, \sigma_2, \sigma_3\}$ - are Pauli matrices. We assume that $U \gg \{J; t\}$. Note that by setting $U_\infty$ we recover the standard canonical t-J model for $n \to 1$ (which we briefly considered in Chapter 5 with respect to the possibility of biholon pairing in the slave-boson formulation of the model in the underdoped case):

$$\hat{H} = -t \sum_{<ij>\sigma} \tilde{c}^+_{i\sigma} \tilde{c}_{j\sigma} + \tilde{J} \sum_{<ij>} \left( \vec{S}_i \vec{S}_j - \frac{1}{4} n_i n_j \right), \quad (13.4.2)$$

with $\tilde{c}_{i\sigma} = c_{i\sigma}(1 - n_{i-\sigma})$ and $\tilde{J} = J + \frac{4t^2}{U} (= J$ for $U \to \infty)$. Note also that the t-J model was derived many years ago by Bulaevskii and coworkers [13.1] to describe the strong-coupling limit of the single-band Hubbard model. The study of this model has become very active in 1990-ties due to Anderson's proposal [13.2] that is was the appropriate model to describe the doped $CuO_2$ planes that are the key ingredients of the high-$T_C$ cuprates. Later on Zhang and Rice [13.3] elucidated the relationship of the t-J model to a multiband Hubbard description with Cu $3d_{x^2-y^2}$ and O $2p_\sigma$ orbitals. The careful numerical investigation of Hybersten and coworkers [13.60] established the parameter values in the mapping of the multiband Hubbard model for the $CuO_2$ planes into a one-band t-J model, namely $J \sim 0.3t$. In the single-band Hubbard model the mapping to a t-J model is valid only in the strong-coupling limit which leads to values $J \ll t$. In a more general model other values of $J/t$ can occur. A lot of work has been done to clarify analytically the relationship between the t-J and multiband Hubbard models, see e.g. [13.61] and reference therein. In this Section we will treat the ratio $J/t$ simply as a parameter to be varied arbitrarily.

Finally let us emphasize that in the canonical form of the t-J model it is convenient to add $-\frac{1}{4} n_i n_j$ to the Heisenberg term $\vec{S}_i \vec{S}_j$ in (13.4.1) and (13.4.2).

In fact the Hamiltonian (13.4.1) of the t-J model with released constraint corresponds to a model with strong onsite repulsion $U$ and small AFM attraction $\sim J$ on the neighboring sites. Effectively we have the Van der Waals interaction potential in this model (see Fig. 13.27 and [13.19]). The bosonic version of the model with Van der Waals interaction was considered in Chapter 6 with respect to the possibility of two-boson pairing.

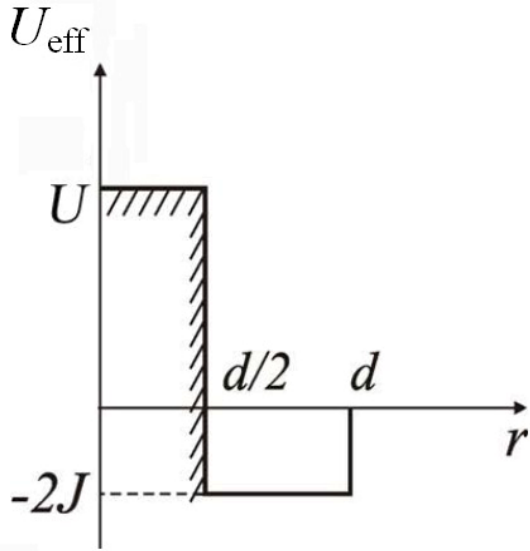

Fig. 13.27. Effective vacuum interaction of the van der Waals type in the 2D isotropic t-J model with released constraint (see also Fig.6.1).

### 13.4.2. SC phase-diagram of the 2D overdoped t-J model.

For small and intermediate electron densities $0 < n_{el} \leq 0.75$ (overdoped case $x \leq 0.25$ for a hole doping) the SC phase-diagram of the 2D t-J model with released constraint has the regions of extended s-wave pairing for $J > 2\,t$ and phase-separation for $J > 3.8\,t$ and $n_{el} \to 0$. For small values of $J/t < 1$ it has the regions of p-wave and $d_{x^2-y^2}$ SC pairing (see [13.18, 13.19] and Fig. 13.28).

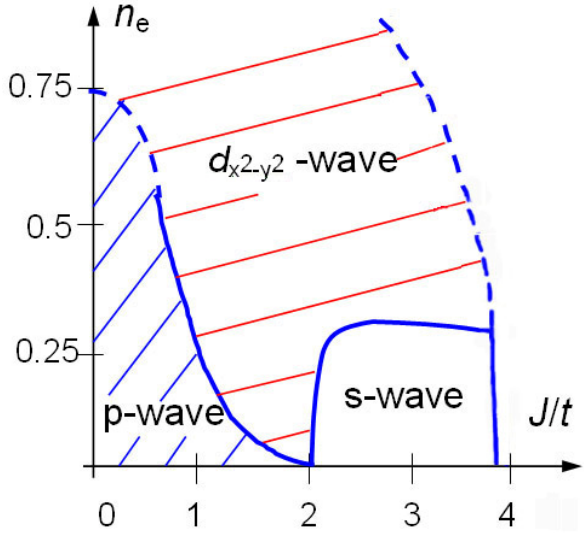

Fig. 13.28. Superconductivity phase-diagram of the 2D t-J model in the overdoped case (for small and intermediate electron densities) [13.19].

### 13.4.3. Extended s-wave pairing for $J > t$ and low electron densities.

At low electron densities and $J > J_C = 2t$ an extended s-wave pairing arises in the 2D t-J model at low electron density (see [13.18]). The superconductive gap for extended s-wave pairing on the 2D square lattice reads (see also Chapter 5):

$$\Delta_S = \Delta_0^S (\cos p_x d + \cos p_y d). \quad (13.4.3)$$

The pair $\Psi$-function is zero for $r \leq d/2$ – in the region of strong Hubbard interaction $U \gg \{J, t\}$ and thus $U\Psi = 0$ in the effective Schroedinger equation. It has a maximum for $r \sim d$ (see Fig.5.1 and Fig. 13.29) which is centered on the neighboring sites. Thus pair $\Psi$-function has region of zero values but does not change sign. For $J > J_C = 2t$ there is a bound state of two electrons with the binding energy:

$$E_b = 8W e^{-\frac{\pi J}{(J - J_{CO})}} \quad (13.4.4)$$

for moderately large $J \geq J_{CO}$ [13.19] and

$$E_b = -J - \frac{20 t^2}{J} \quad (13.4.5)$$

for large $J \gg t$ [13.18].

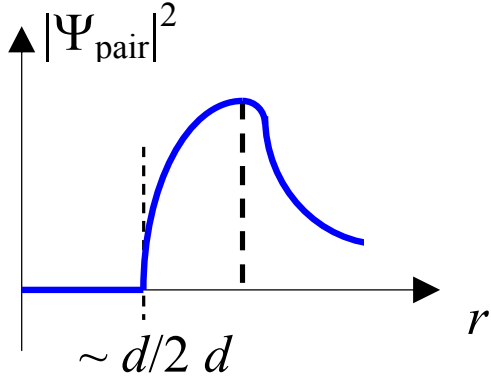

Fig. 13.29. Pair Ψ-function squared for the extended s-wave pairing in the 2D t-J model (see also Fig.5.1 in Chapter 5).

In the BCS-case for $|E_b| < \varepsilon_F$ and extended Cooper pairing the mean-field superconductive critical temperature is given by famous Miyake formula [13.62] (see Chapters 5 and 14):

$$T_{Cs} \sim \sqrt{2\varepsilon_F |E_b|} \qquad (13.4.6)$$

In the BEC-case for $|E_b| > \varepsilon_F$ and local pairing we have two characteristic temperatures:

$$T_* \sim \frac{|E_b|}{\ln(|E_b|/\varepsilon_F)} - \qquad (13.4.7)$$

- Saha crossover temperature [13.63] which describes creation of local pairs (dimers) and superconductive critical temperature:

$$T_{CS} = \frac{\varepsilon_F}{4\ln\ln(|E_b|/\varepsilon_F)} \qquad (13.4.8)$$

given by Fisher, Hohenberg formula [13.64] for slightly non-ideal 2D Bose-gas with repulsion between local pairs (dimers) (see also Popov [13.65]). The more detailed discussion of the BCS-BEC crossover
[13.66-13.68] in 2D attractive Fermi-gas is presented in Chapters 6 and 14.

13.4.4. Phase-separation at large J/t and low electron density.

The energy of BEC-phase becomes negative at large $J/t$ – we have here a liquid of local pairs (dimers) with an energy:

$$E_{BEC} - E_N = -N\frac{|E_b|}{2} < 0, \qquad (13.4.9)$$

where $N$ is a number of particles.

If we further increase the J/t-ratio at low electron densities, than the formation of quartets [13.19, 13.69, 13.16] or larger complexes will become energetically beneficial on the 2D square lattice. But as it was shown by Emery, Kivelson and Lin [13.18], the liquid phase for $J > J_{P.S}$ = 3.8 t (which is formed earlier then the threshold for quartet formation) becomes unstable towards total phase-separation (see also Chapter 17) on two large clusters: PM-cluster with electron density $n_{el} \to 0$ and AFM-cluster with $n_{el} \to 1$. The threshold value for the total phase-separation $J_{P.S}$ = 3.8 t can be defined from a simple estimate (see also numerical calculations of Dagotto et al. [13.70]):

$$\frac{E_{AFM} - E_N}{N} = -\frac{1.18J}{2} \leq \frac{E_{BEC} - E_N}{N} = -\frac{|E_b|}{2}, \qquad (13.4.10)$$

where $1.18J$ is AFM-energy per bond for the 2D square lattice.

### 13.4.5. p-wave pairing for $J < t$ and low electron densities.

For small values of $J/t$ and low electron densities the triplet p-wave pairing, governed by Kohn-Luttinger mechanism [13.20, 13.71, 13.72] corresponds to a leading SC-instability in the system below the critical temperature:

$$T_{Cp} \sim \varepsilon_F e^{-\frac{1}{6.1 f_0^3}} \qquad (13.4.11)$$

(see Chapter 11 for more details). The p-wave SC gap for the 2D square lattice reads:

$$\Delta_p = \Delta_{0p}(\cos p_x d + i \sin p_y d). \qquad (13.4.12)$$

In the case of 2D t-J model with released constraint in (13.4.11) the coupling constant $f_0$ for $J < J_{CO} = 2t$ is given by [13.19]:

$$f_0 = \frac{1}{\ln\frac{4W}{\varepsilon_F} + \frac{\pi J}{J_{CO} - J}} \qquad (13.4.13)$$

Thus for $J/t \to 0$ the coupling constant $f_0 = \left(\ln\frac{4W}{\varepsilon_F}\right)^{-1}$ as in the 2D Hubbard model [13.73]. It is possible to demonstrate by direct comparison of the critical temperatures in different channels (see Fig. 13.35) that p-wave pairing is dominant for $J < t$ and low electron densities $n_{el} = 2\varepsilon_F/W \ll 1$.

### 13.4.6. d-wave pairing in the overdoped 2D t-J model.

For $n_{el} \geq (0.6\text{-}0.7)$ and not very small ratio of $J/t \sim (½ \div ⅓)$, which are just typical values for high-$T_C$ materials, d-wave pairing becomes dominant over p-wave pairing in the 2D t-J model with released constraint. The equation for critical temperature in $d_{x^2-y^2}$-channel reads (see [13.19]) in the weak-coupling case $J/t < 1$:

$$1 = Jd^2 \iint \frac{dp_x}{2\pi} \frac{dp_y}{2\pi} \varphi_d^2 \frac{\text{th}\left(\frac{\varepsilon_p - \mu}{2T_C}\right)}{2(\varepsilon_p - \mu)}, \qquad (13.4.14)$$

where $J$ is just AFM attractive interaction, $\phi_d = (\cos p_x d - \cos p_y d)$ is an eigenfunction for $d_{x^2-y^2}$-pairing on the 2D square lattice, $\varepsilon_p - \mu = -2t(\cos p_x d + \cos p_y d) - \mu$ is the uncorrelated quasiparticle spectrum in 2D.
As a result we get for d-wave pairing [13.19, 13.49]:

$$T_{Cd} \sim \varepsilon_F e^{-\frac{\pi t}{2 J n_{el}^2}} \qquad (13.4.15)$$

Extrapolation of these results on $J/t \sim (½ \div ⅓)$ and $n_{el} \sim 0.85$ ($x \sim 0.15$ - optimal doping) yields the rough estimate $T_{Cd} \sim \varepsilon_F e^{-5} \sim 10^2$ K for $\varepsilon_F \sim 10^4$ which is quite reasonable for cuprates.

### 13.4.7. d-wave pairing at small hole densities $x = (1 - n_{el}) \ll 1$.

In the opposite case of small hole densities $x = (1 - n_{el}) \ll 1$ the similar to (13.4.14) equation for $T_C$ with the spin-polaronic spectrum $\varepsilon(p)$ was derived by Plakida's group [13.26] using diagrammatic technique for the Hubbard operators [13.74, 13.75]. In the weak-coupling BCS-case for $T_C < E_F(x)$ the critical temperature in the paramagnetic region reads [13.26]:

$$T_C^d \sim \sqrt{W E_F(x)} e^{-\frac{1}{\lambda}} \qquad (13.4.16)$$

where $\lambda \sim J N_0(x) \sim 0.3$ is a coupling constant, $E_F(x)$ is Fermi-energy, $N_0(x)$ is an averaged density of states. $T_C$ is maximal at optimal doping $x \sim x_{opt} \sim 0.15$ where $E_F(x_{opt}) \sim W/2$ and where we effectively have a crossover from a small hole-like Fermi-surface to a large electronic one.

In maximum again $T_C^d \sim 10^2$ K. Note that Plakida et al., also considered generalized one-band 2D t-J model derived from multiband Hubbard model (or two-band Emery model [13.76]) when we neglect the interband Hubbard repulsion between d- and p-orbitals ($U_{dp} = 0$). In this case the local constraint (see (13.4.2)) is also not very important (as in the Kagan, Rice approach [13.19]) and we can neglect also the kinematical interaction of Zaitsev et al. [13.77].

13.4.8. Possible bosonic region of the phase-diagram of the 2D t-J model in the underdoped case.

In the extreme underdoped case very close to half-filling for $x \ll x_{opt}$ we have the physics of pseudogap at $T_C \leq T \leq T_*$ (see also Chapter 14) and a bosonic-type Uemura plot for $T_C$ ($T_C(x) \sim x$) [13.78]. If we assume the bosonic character of the pseudogap (connected with SC-fluctuations of preformed pairs, and not with AFM-fluctuations), then we could expect the formation of local pairs consisting of two spin-polarons at some higher temperatures $T_* \sim |E_b|$ (their binding energy) and BEC of local pairs at lower temperatures $T_C \leq E_F(x) \leq T_*$. Note that in this region of the phase-diagram we have small hole Fermi-surface with $E_F(x) \sim Jx$ according to Lee et al. [13.45]. In this limiting case our philosophy, however, is more close to the ideas of Laughlin et al. [13.21, 13.22] on spin-charge confinement than to the philosophy of Anderson [13.2] and Lee [13.43, 13.44] on spin-charge separation in the 2D t-J model (see also Larkin et al., [13.89] on spin-charge binding in the t-J model). Note that there is a crucial difference between spin-charge binding and spin-charge confinement. While in the first case we have spin-charge binding at low temperatures and spin-charge separation at high temperatures, in the second case we have spin-charge confinement everytime (at arbitrary high temperatures). Let us stress that Laughlin [13.21, 13.22] assumed the spin-charge confinement in the strongly interacting Fermi-Bose mixture of spinons and holons at small hole density in analogy with the confinement in quark-gluon plasma in QCD [13.50-13.54]. As we already mentioned in the introduction to this Chapter, the spin-charge confinement leads to the creation of composite holes [13.21, 13.22] (or spin-polarons [13.23, 13.24] or strings [13.1, 13.17, 13.25]). The basic results here are connected with the ideas of Bulaevskii, Nagaev, Khomskii [13.1] and Brinkman-Rice [13.17] on AFM-string for a hole motion in 2D AFM-background of spins $S = \frac{1}{2}$).

13.4.9. String-like solution for a composite hole.

The illustration of the formation of the confinement potential (of the linear trace of frustrated spins which accompany a hole motion in 2D AFM-background of spins $S = \frac{1}{2}$) is presented on Fig. 13.30.

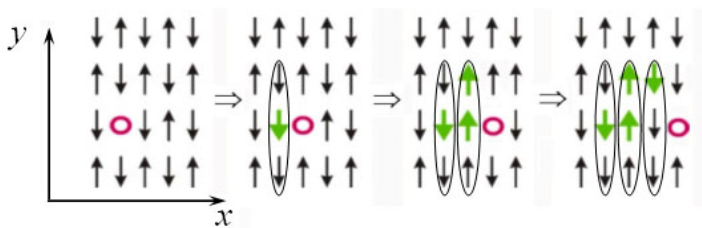

Fig. 13.30. Formation of the string for a motion of a hole along the horizontal x-axis in the right-hand side in 2D AFM-background of spins $S = \frac{1}{2}$ on the square lattice.

In a simple picture the $\Psi$-function of a string is a solution of a Shroedinger equation with linear potential $V(r) = \frac{zJS^2}{2} r$ for a spinon-holon interaction ($z$ is the number of nearest neighbors ($z = 4$) on the square lattice):

$$-\frac{\hbar^2}{2m}\Delta\Psi + \frac{zJS^2}{2} r\Psi = E\Psi, \qquad (13.4.17)$$

The solution of this expression is given by [13.79] $\Psi \sim Ai(r)$ for Airy function. The effective radius of a string-oscillator [13.1] does not depend upon dimensionality for $D = 2$ and 3 and yields:

$$r_0 \sim \left(\frac{t}{zJS^2}\right)^{1/3}. \qquad (13.4.18)$$

As a result the energy of a string:

$$E_0 \sim -\frac{Zt}{\sqrt{Z-1}} + \left(zJS^2\right)^{2/3} t^{1/3}, \qquad (13.4.19)$$

where the bottom of the band also changes for a string motion [13.31].

An account of the quantum fluctuations connected with the term $J(S_i^+ S_j^- + S_i^- S_j^+)$ in the 2D t-J model leads to the dispersion of composite hole with a spectrum (see [13.31])

$$E_h(q) = E_0 + J(\cos q_x d + \cos q_y d)^2 \qquad (13.4.20)$$

(here we neglect the difficulties connected with the so-called Trugman paths which could destroy a string after several traversing of elementary plaquette assuming as usual that their statistical weight is small [13.80]).

The Green-function of a composite hole has a simple one-pole structure of the type (see Eq (13.1.7) and also Lee et al. [13.43-13.45]):

$$G(\omega,\vec{q}) \sim \frac{J/t}{\omega - E_0 - Jq^2 + \mu + io} + G_{incoh}(\omega,\vec{q}).$$

13.4.10. The two-particle problem for composite holes. Possibility of BCS-BEC crossover in the d-wave channel.

Residual interaction of the two composite holes for a small hole concentration $x \ll 1$ has a dipole-dipole character according to hydrodynamic approach of Shraiman, Siggia [13.25]:

$$V(r) \sim \frac{\lambda}{r^2} \qquad (13.4.21)$$

It was shown by Belinicher group [13.23, 13.24] that this interaction can lead to a shallow bound state of the two composite holes (two spin-polarons) in the $d_{x^2-y^2}$-wave channel. It is quite appealing to consider $T_C$ versus $x$ dependence for strongly underdoped high-$T_C$ superconductors as the BCS-BEC crossover for the pairing of two composite holes (two spin-polarons) in the d-wave channel [13.16].

Note that if we solve the two-particle problem for composite holes (two string-oscillators) interacting via dipole-dipole potential we will find according to Belinicher et al. [13.23, 13.24] the binding energy for $d_{x^2-y^2}$-pairing:

$$|E_b| \sim 0.02\, t \sim T_* \qquad (13.4.22)$$

In the same time the BEC critical temperature for small hole concentration reads:

$$T_C^{BEC} \sim Jx < T_* \qquad (13.4.23)$$

where $J \sim (0.3 \div 0.5)\, t$ and effective mass of a pair $m^* \sim 1/J$.

In the opposite limit of larger hole concentration, as we already mentioned, we have BCS-type $d_{x^2-y^2}$-pairing described by Kagan, Rice [13.19] and Plakida [13.26].

Concluding this Chapter, note once more that while spin-chains and three-leg ladders are grossly described by the physics of spin-charge separation, the two-leg ladders and possibly underdoped high-$T_C$ materials are more close to the ideas of the physics of spin-charge confinement. In the same time strongly overdoped 2D cuprates are well described by more standard Landau Fermi-liquid picture and show the tendency towards superconductive instabilities for p-wave and $d_{x^2-y^2}$-pairing at low $J/t$-values as well as towards extended s-wave pairing and total phase-separation at large $J/t$-values and low electron densities.

The rough extrapolation of the low electron density results on optimal doping yields for parameter values typical for cuprates $J/t \sim (½ \div ⅓)$ and $n_{el} \sim 0.85$ the reasonable temperatures for d-wave pairing ($T_{Cd} \sim 10^2$ K).

The physics of the strongly underdoped t-J model could be possibly described by the scenario of the BCS-BEC crossover for the pairing of two composite holes (two spin-polarons or two AFM strings) in the $d_{x^2-y^2}$-wave channel.

Reference list to Chapter 13.